\documentclass[
superscriptaddress,
amsmath,amssymb,
aps,
prl,
twocolumn
]{revtex4-2}
\usepackage{hyperref}
\usepackage{graphicx}
\usepackage[export]{adjustbox}

\newcommand{\rme}{{\rm e}}

\newcommand{\rmi}{{\rm i}}

\begin{document}
\title{Quantum jumps in driven-dissipative disordered many-body systems}
\author{Sparsh Gupta}
\email{sparsh.gupta@icts.res.in}
\affiliation{International Centre for Theoretical Sciences, Tata Institute of
Fundamental Research, 560089 Bangalore, India}
\author{Hari Kumar Yadalam}
 \email{hyadalam@uci.edu}
\affiliation{Laboratoire de Physique de l'\'Ecole Normale Sup\'erieure, ENS, Universit\'e PSL, CNRS, Sorbonne Universit\'e, Universit\'e Paris Cit\'e, F-75005 Paris,  France}
\affiliation{International Centre for Theoretical Sciences, Tata Institute of
Fundamental Research, 560089 Bangalore, India}
\affiliation{Department of Chemistry, University of California, Irvine, CA 92614, USA}
\affiliation{Department of Physics and Astronomy, University of California, Irvine, CA 92614, USA}
\author{Manas Kulkarni}
 \email{manas.kulkarni@icts.res.in}
\affiliation{International Centre for Theoretical Sciences, Tata Institute of Fundamental Research, 560089 Bangalore, India}
\author{Camille Aron}
 \email{aron@ens.fr}
   \affiliation{Institute of Physics, \'Ecole Polytechnique F\'ed\'erale de Lausanne (EPFL), CH-1015 Lausanne, Switzerland}
\affiliation{Laboratoire de Physique de l'\'Ecole Normale Sup\'erieure, ENS, Universit\'e PSL, CNRS, Sorbonne Universit\'e, Universit\'e Paris Cit\'e, F-75005 Paris,  France}
\date{\today}

\begin{abstract}
We discuss how quantum jumps affect localized regimes in driven-dissipative disordered many-body systems featuring a localization transition. We introduce a deformation of the Lindblad master equation that interpolates between the standard Lindblad and the no-jump non-Hermitian dynamics of open quantum systems. As a platform, we use a disordered chain of hard-core bosons with nearest-neighbor interactions and subject to incoherent drive and dissipation at alternate sites. We probe both the statistics of complex eigenvalues of the deformed Liouvillian and dynamical observables of physical relevance. We show that reducing the number of quantum jumps, achievable through realistic postselection protocols, can promote the emergence of the localized phase. Our findings are based on exact diagonalization and time-dependent matrix-product states techniques.

\end{abstract}


\maketitle
\paragraph*{Introduction.}
Sufficiently strong disorder can markedly hinder the dynamics of many-body systems. Quantum many-body localized regimes~\cite{Fleishman1980Interactions,Altshuler1997Quasiparticle,Basko2006Metal,
Nandkishore2015Many,AletLaflorencie2018}, wherein transport is completely arrested, have attracted considerable attention because their inability to thermalize evades the foundations of statistical mechanics~\cite{Deutsch1991Quantum,Srednicki1994Chaos,Rigol.2008,QChaos_DAlessio.2016,Deutsch2018Eigenstate}.
The delocalized and localized regimes, at respectively weak and strong disorder, are commonly probed by means of spectral~\cite{Oganesyan2007Localization,Pal2010Many} and dynamical properties~\cite{Abanin2017Recent,Abanin2019Colloquium}.
The eigenvalue statistics are generically expected to transition from Hermitian random-matrix to one-dimensional Poisson statistics as the strength of the disorder increases. The transport and information-spreading properties at strong disorder are expected to display signs of non-ergodicity, such as dependence on initial conditions or logarithmic growth of entanglement~ \cite{EEgrowth_Bardarson.2012,EEgrowth_Serbyn.2013}. Various experiments demonstrating localization transitions have been successfully conducted across different quantum many-body platforms~\cite{Schreiber2015Observation,Smith2016Many,choi2016exploring,
roushan2017spectroscopic,Fan2018,Cappellaro2018,lukin2019probing,Gong2021Experimental}.

\begin{figure}
\centering
\includegraphics[width=5.8cm]{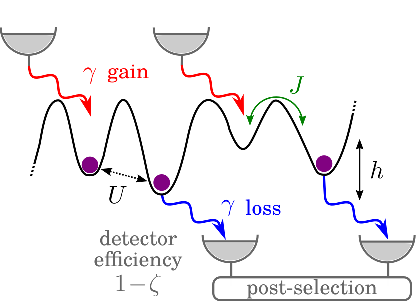}
\caption{
Sketch of the disordered gain-loss model, see the Hamiltonian $H$ in Eq.~(\ref{eq:H}) and proposed protocol to implement the $\zeta$-deformed Liouvillian $\mathcal{L}_\zeta$ in Eq.~(\ref{eq:Liouvillian_zeta_main}). Both gain and loss events are monitored by means of realistic detectors with efficiency $0 \leq 1-\zeta \leq 1$.
Here, the postselection interpretation consists in selecting those monitored trajectories with no jump. See the Supplementary Material~\cite{SM} for details and an alternative protocol.
}
\label{fig:sketch}
\end{figure}

The inevitable presence of an environment is expected to destabilize localized regimes, confining their existence to intermediate timescales before complete thermalization with the environment~\cite{Fischer2016Dynamics,Medvedyeva2016Influence,Everest2017Role,Vakulchyk2018Signatures,EPL2018Dynamics,Lenarcic2020Critical,Kos2021Thermalization,Mak2023Statistics,Luschen2017Signatures,Rubio2019Many}.
However, it has recently been demonstrated that non-equilibrium environments could sustain localization~\cite{Hamazaki2019Non,Panda2020Entanglement,Zhai2020Many,Yamamoto2022Universal,Dai2022Dynamical,Suthar2022Non,Ghosh2022Spectral}, sparking renewed interest in many-body localization in driven-dissipative settings.
These are often described by non-Hermitian Hamiltonians, where the non-Hermiticity mimics the hybridization with reservoirs and can be interpreted in terms of postselection protocols~\cite{Hamazaki2019Non,Ueda2020Non,Yamamoto2022Universal,Hamazaki2019Non,Panda2020Entanglement,Zhai2020Many,Yamamoto2022Universal,Dai2022Dynamical,Suthar2022Non,Ghosh2022Spectral}.
A natural approach that does not rely on postselection interpretation and which can systematically incorporate the effects of Markovian environments is the standard Lindblad quantum master equation approach~\cite{breuer2002theory,carmichael2009open,daley2014quantum}. In this approach, the environment contributes to two types of processes: the ones that can be absorbed in a non-Hermitian Hamiltonian description, and others that can be interpreted in terms of quantum jumps~\cite{dalibard1992wave,molmer1993monte,dum1992monte,molmer1996monte,plenio1998the,carmichael2009open,daley2014quantum}.
Recently, frameworks bridging these two approaches have been developed~\cite{Konstantin2014Comparison,Minganti2019Quantum,Minganti2020Hybrid}. They rely on suitable deformations of the standard Lindblad equation and can be experimentally motivated~\cite{Bergquist1986Observation,Sauter1986Observation,Minev2019Catch,Chen2021Quantum}.

In this Letter, we question the precise role of quantum jumps on the fate of localized regimes by working with a disordered one-dimensional many-body system coupled to a gain-loss environment. See the schematic in Fig.~\ref{fig:sketch}. 
To that end, we first introduce a specific deformation of the standard Lindblad master equation that involves a parameter $\zeta \in [0, 1]$ dialing the strength of quantum jump terms all the way from the non-Hermitian Hamiltonian to the standard Lindblad description.
Our analysis of the influence of quantum jumps on the complex spectrum as well as on the dynamics of the $\zeta$-deformed Liouvillian demonstrates that fewer quantum jumps can result in the emergence of localization at lower disorder strengths.
To put it another way, postselection can promote localization.
We emphasize that this is not only a formal construction but it can also find an experimental realization with realistic faulty detectors.

\paragraph*{$\zeta$-deformed theory.}
The Markovian evolution of open quantum systems is generically described by the Lindblad equation
$\partial_t\rho(t) = \mathcal{L} \rho(t)$,
with the Liouvillian
$ \mathcal{L} \mathbb{\star} := -\rmi\left[H,\mathbb{\star}\right]+\sum_{\alpha}^{}\left[O_\alpha^{} \mathbb{\star} O_\alpha^\dagger -\left\{O^\dagger_\alpha O_\alpha^{},\mathbb{\star}\right\}/2\right]$,
where $H$ is the Hermitian Hamiltonian of the system and the $O_\alpha$'s with $\alpha = 1, \ldots, M$ are the jump operators in the $M$ dissipative channels. 
This Lindblad evolution can be unraveled into a quantum-jump trajectory ensemble as
$ \rho(t) = \sum_{n=0}^{\infty} \rho_{n}(t)$
where $\rho_{n}(t)$ is the conditional density matrix of the system subjected to precisely $n$ quantum jumps until time $t$~\cite{dalibard1992wave,molmer1993monte,dum1992monte,molmer1996monte,plenio1998the,carmichael2009open,daley2014quantum}.
Let us introduce a weight $\zeta \in [0,1]$ to each jump in a quantum trajectory. In analogy to the familiar terminology of the grand-canonical ensemble, we coin it the “quantum-jump fugacity”.
This defines a $\zeta$-deformed ensemble where the density matrix 
$ \rho_{\zeta}(t) = {\sum_{n=0}^{\infty} \zeta^{n} \rho_n(t)}/{\sum_{n=0}^{\infty} \zeta^{n}{\rm Tr}\left[\rho_n(t)\right]}$
evolves according to the following $\zeta$-deformed Lindblad master equation,
\begin{align}\label{eq:Lindblad_zeta_main}
\partial_t  \rho_\zeta(t)&= \Big{(}
 \mathcal{L}_{\zeta}-\mathrm{Tr}\left[\mathcal{L}_{\zeta}\rho_{\zeta}(t)\right]\Big{)} \rho_{\zeta}(t)\,,
\end{align}
where the $\zeta$-deformed Liouvillian is given by
\begin{align}\label{eq:Liouvillian_zeta_main}
 \mathcal{L}_{\zeta}\star  := -\rmi\left[H,\star\right]+\sum_{\alpha=1}^{M}\left[\zeta O_\alpha\star O_\alpha^\dagger -  \frac{1}{2}\left\{O_\alpha^\dagger O_\alpha,\star\right\}\right].
\end{align}
The systematic and consistent construction of such a theory is detailed in the Supplementary Material~\cite{SM}.
The standard Lindblad equation is recovered in the limit $\zeta = 1$ whereas the limit $\zeta = 0$ corresponds to an evolution generated by the non-Hermitian Hamiltonian $\tilde H = H -  \frac{\rmi}{2} \sum_{\alpha=1}^M O_\alpha^\dagger O_\alpha$. 
The subscript $\zeta$ in $\rho_\zeta(t)$ is to distinguish between the results of the deformed theory and of the standard Lindblad evolution. The initial condition is given by $\rho_\zeta(0) = \rho(0)$ and the evolution in Eq.~(\ref{eq:Lindblad_zeta_main}) is completely positive, Hermiticity and trace preserving.
The latter is ensured by the non-linear trace term.
The observables predicted from Eq.~(\ref{eq:Lindblad_zeta_main}) can be experimentally measured by postselection protocols. In Fig.~\ref{fig:sketch}, we depict a possible protocol where both the quantum jumps due to gain and loss processes are monitored by means of detectors characterized by an efficiency $1-\zeta$, \textit{i.e.} the error rate of returning a no-click result when a jump occurred is $\zeta$. Here, the postselection protocol consists of discarding those trajectories where one or more jumps were monitored.
Decreasing the efficiency of the detectors increases the average number of quantum jumps in the post-selected dynamics.
We discuss an alternative protocol in the Supplementary Material~[\onlinecite{SM}].
We note that generalized Lindblad equations of the type of Eq.~(\ref{eq:Lindblad_zeta_main}) appear in the studies of full-counting statistics, where they are referred to as tilted or twisted master equations \cite{Esposito2009Nonequilibrium,Garrahan2010Thermodynamics,Yadalam2022Counting}.

\begin{figure}
\centering
\includegraphics[width=4.2cm,valign=M]{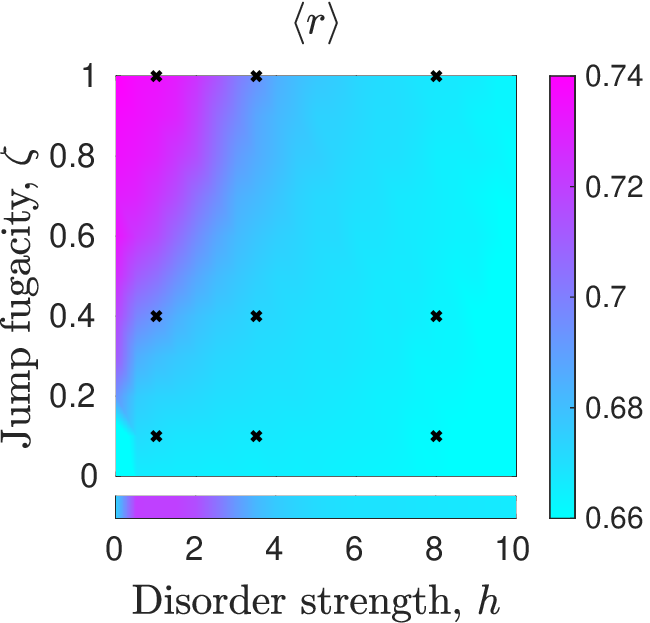} 
\includegraphics[width=4.2cm,valign=M]{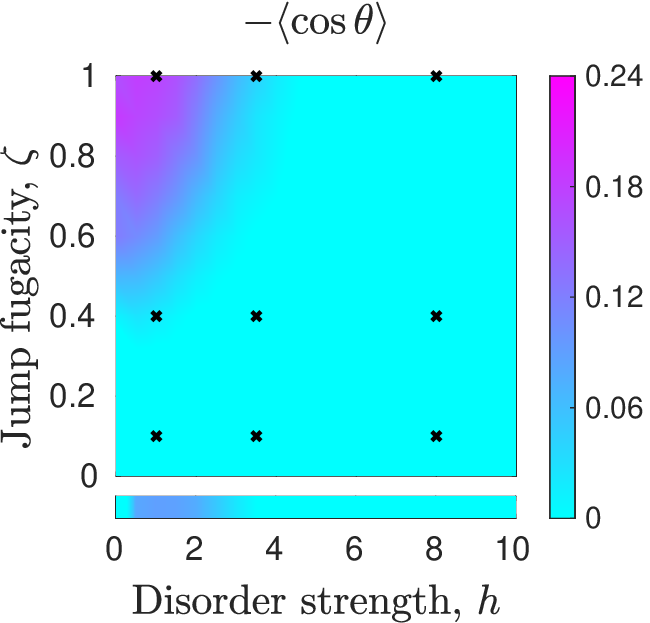}
\caption{
Complex spacing ratios of the spectrum of $\mathcal{L_\zeta}$ in Eq.~(\ref{eq:Liouvillian_zeta_main}): (left) $\langle r \rangle$  and (right) $-\langle \cos \theta \rangle$ computed by exact diagonalization of a system of $L=8$ sites, in the zero charge sector of the weak $U(1)$ symmetry, and averaged over $160$ disorder samples. The bottom strips are obtained using the non-Hermitian Hamiltonian $\tilde H$ (half-filling sector) in Eq.~(\ref{eq:H_NH}) for system size $L=16$ averaged over $160$ disorder samples. 
The nine black crosses in each panel correspond to the parameters at which the densities of complex spacing ratios are presented in Fig.~\ref{fig:NNN_spacingratio}.
}
\label{fig:avg_level_spacing_ratio}
\end{figure}

\begin{figure}
\centering
\includegraphics[width=7cm]{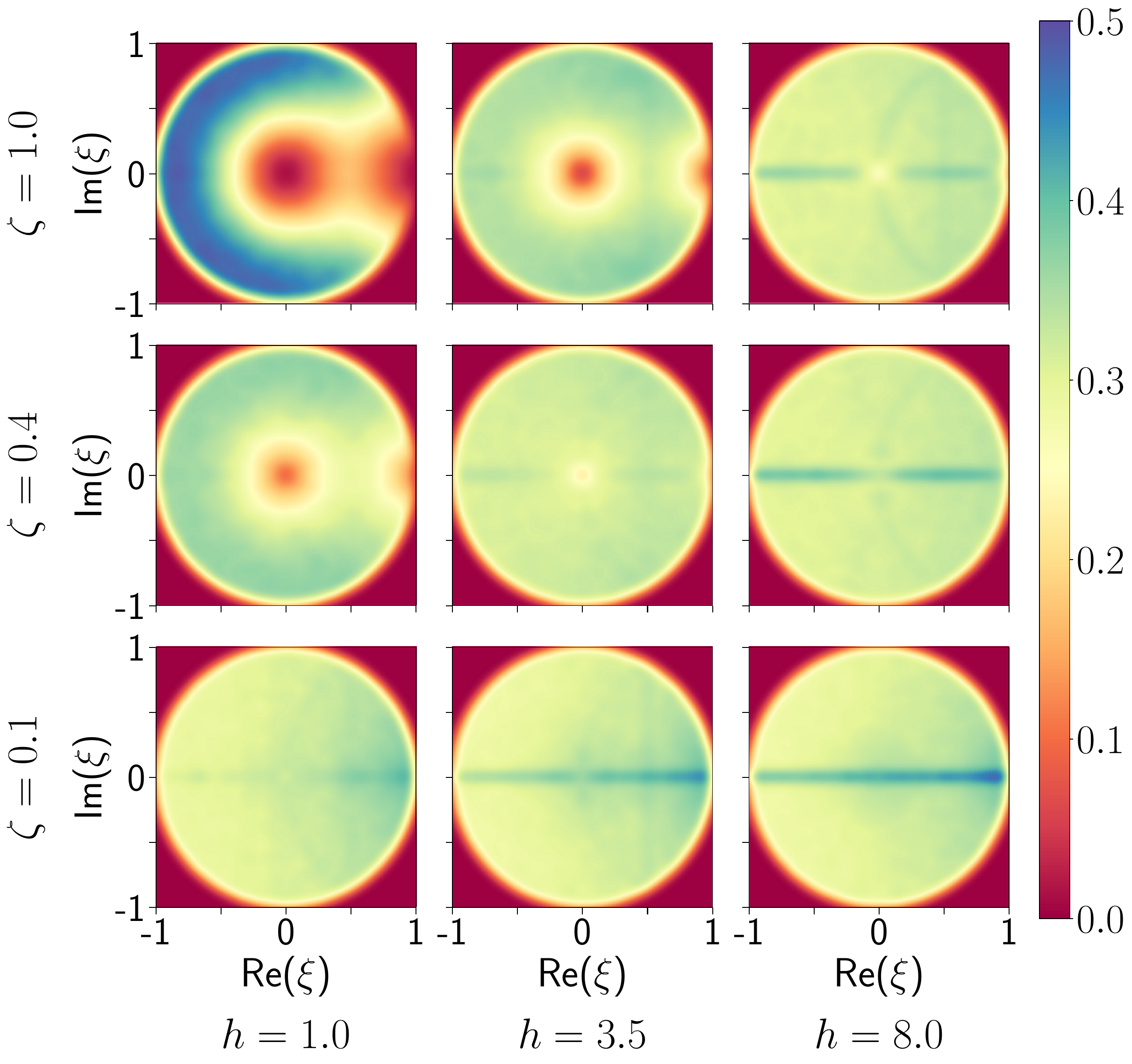}
\caption{Density of complex spacing ratio defined in Eq.~(\ref{eq:csr}) for representative values of disorder strength $h$ and quantum-jump fugacity $\zeta$ for a system of $L=8$ sites in the  zero charge sector of the weak $U(1)$ symmetry, and averaged over $160$ disorder samples. The isotropy associated with the localized phase increases with decreasing $\zeta$ or increasing $h$.
}
\label{fig:NNN_spacingratio}
\end{figure}

\paragraph*{Disordered gain-loss model.}
To understand the role of quantum jumps on the localized-delocalized transition in non-Hermitian many-body systems, we consider a disordered gain-loss model defined by the following Hamiltonian (see Fig.~\ref{fig:sketch})
\begin{align} \label{eq:H}
 H =  \!\sum_{i=1}^{L} h_i n_i - J \!\sum_{i=1}^{L-1}  \!\left(b_i^\dagger b_{i+1} \!+\! \mathrm{H.c.}\!\right) + U \! \sum_{i=1}^{L-1}  n_i n_{i+1},
\end{align} 
with $n_i = b_i^\dagger b_i$, and by the onsite jump operators
\begin{align}
O_i &=\begin{cases} \sqrt{2\gamma} \ b_i^\dagger & \text{if} \ i \ \text{is odd}\\
      \sqrt{2\gamma} \ b_i & \text{if} \ i \ \text{is even}\,.
     \end{cases}
\end{align}
The $b_i^\dagger$'s and $b_i$'s, $i = 1,\, \ldots,\, L$, are onsite creation and annihilation operators of hard-core bosons living on a one-dimensional lattice with $L$ sites and open boundary conditions. The Hamiltonian in Eq.~(\ref{eq:H}) is $U(1)$-symmetric, \textit{i.e.} it conserves the total number of particles $N = \sum_{i=1}^L n_i$.  $h_i$ are independent random energy levels uniformly distributed in the interval $\left[-h,h\right]$. $J$ is the inter-site hopping amplitude. $U$ is the inter-site interaction which we set to $U=2J$ for the Hamiltonian to be equivalent to the disordered Heisenberg spin chain which has been extensively studied in the context of Hermitian many-body localization~\cite{Oganesyan2007Localization,Pal2010Many,Abanin2017Recent,Abanin2019Colloquium}.
Its transition was found around $h^\star \approx 7J$ in our conventions.
$\gamma$ sets the rates of both the incoherent gain and loss occurring at alternating sites, which we set as $\gamma = 0.1 J$ throughout this work.
We choose $J$ as the unit of energy and, therefore, we set $J=1$.
The corresponding $\zeta$-deformed Liouvillian $\mathcal{L}_\zeta$ in Eq.~(\ref{eq:Liouvillian_zeta_main}) has a weak $U(1)$ symmetry that corresponds to the conservation of the particle number difference between the bra and ket sides of the states upon acting with $\mathcal{L}_\zeta$, see details in the Supplementary Material~[\onlinecite{SM}].

In the limit of $\zeta = 0$, there is an additional weak $U(1)$ symmetry of $\mathcal{L}_0$  that corresponds to conserving the particle number associated with the bra and ket independently.
Moreover, the $\zeta = 0$ dynamics boils down to that of the non-Hermitian gain-loss Hamiltonian
\begin{align} \label{eq:H_NH}
\tilde{H} = H - \rmi \gamma \sum_{i=1}^L (-1)^{i} b_i^\dagger b_i\,, \qquad \zeta = 0\,,
\end{align}
recently studied in Refs.~\cite{Hamazaki2019Non,Lucas2020Complex,Ghosh2022Spectral}.
$\tilde{H}$ also conserves the total number of particles. It displays a non-Hermitian many-body localization transition at $h^\star \approx 4.2$ manifesting itself as a crossover between AI$^{\dagger}$ non-Hermitian random-matrix (weak disorder) and two-dimensional Poisson ensembles (strong disorder).
Here, we explore this physics both from spectral and dynamical points of view in the general $\zeta$-deformed Lindbladian framework that captures the effect of quantum jumps in a controllable fashion.

\paragraph*{Spectral signatures.}
The Liouvillian $\mathcal{L}_\zeta$ in Eq.~(\ref{eq:Liouvillian_zeta_main}) is a non-Hermitian operator and we analyze its complex spectrum by means of exact diagonalization. We specifically compute the statistics of the complex spacing ratio~\cite{Lucas2020Complex} defined for each eigenvalue $z$ as 
\begin{equation}\label{eq:csr}
\xi = \frac{z^{\rm NN}-z}{z^{\rm NNN}-z} = r \, \rme^{\rmi \theta}\,,
\end{equation}
where $z^{\rm NN}$ and $z^{\rm NNN}$ are the nearest and the next-nearest neighbor eigenvalues to $z$ (in euclidean distance), respectively. $r$ and $\theta$ are respectively the norm and the argument of $\xi$.
Note that the non-linear trace term in $\mathcal{L}_\zeta$ simply adds a constant shift to the spectrum and is therefore inconsequential to level-spacing statistics.
The statistics of $\xi$ are indicative of the chaotic or regular nature of complex-valued spectra and have been studied in the context of non-Hermitian interacting disordered Hamiltonians~\cite{Lucas2020Complex,csr_syk2022,Suthar2022Non} and open quantum systems described by standard Lindblad evolutions~\cite{Lucas2020Complex,wang2020,garcia2022, Hamazaki2022Lindbladian,yusipov2022,Mahaveer2022Dissipative,ferrari2023steadystate}.
For chaotic systems, the eigenvalues experience level repulsion resulting in a vanishing complex spacing ratio distribution at small $r$ and an anisotropic angular pattern.
The distributions of $r$ and $\theta$ are generically dictated by Ginibre random matrix ensembles, and their averages take the value $\langle r \rangle \approx 0.738$ and $-\langle  \cos\theta \rangle  \approx 0.244$.
On the other hand, for uncorrelated energy levels, the complex spacing ratio is uniformly distributed inside a unit circle~\cite{Lucas2020Complex} with $\langle r \rangle= 2/3$ and $-\langle \cos \theta \rangle = 0$.

In Fig.~\ref{fig:avg_level_spacing_ratio}, we present $\langle r\rangle$ and $-\langle \cos \theta \rangle$ as a function of both the disorder strength $h$ and the quantum-jump fugacity $\zeta$. The results are obtained from the zero charge sector of the weak $U(1)$ symmetry for a system of $L=8$ sites and after averaging over $160$ disorder samples.
For the standard Lindblad evolution at $\zeta = 1$, we find a clear transition between random matrix predictions at weak disorder and two-dimensional Poisson predictions at strong disorder.
When $\zeta$ is reduced, the location of this transition is shifted to lower disorder strengths: reducing the number of quantum jumps facilitates the emergence of localization.
In the $\zeta=0$ case, the additional weak $U(1)$ symmetry is responsible for spurious statistics which are known to produce deceitful level attraction between eigenvalues of different symmetry sectors.
We attribute the apparent loss of a delocalized phase in the vicinity of $\zeta =0$ to a remnant of this extra symmetry.
To circumvent this situation at $\zeta=0$, one should resort to analyzing the spectrum of the non-Hermitian Hamiltonian~\cite{Hamazaki2019Non,Lucas2020Complex,Ghosh2022Spectral} in Eq.~(\ref{eq:H_NH}). The corresponding results are presented in the strips of Fig.~\ref{fig:avg_level_spacing_ratio}.
In Fig.~\ref{fig:NNN_spacingratio}, we illustrate further these effects of disorder and quantum-jump fugacity by presenting representative plots of the density of complex spacing ratios $\xi$ in Eq.~(\ref{eq:csr}) for different values of $h$ and $\zeta$.
At strong disorder and weak fugacity, we find that this distribution is isotropic and homogeneous within the unit circle, which is a hallmark of integrable systems. On the other hand, for weak disorder and  large fugacity, the distribution is found to be anisotropic and inhomogeneous, which is expected for chaotic systems.

\begin{figure}
\centering
\includegraphics[width=4.2cm,valign=M]{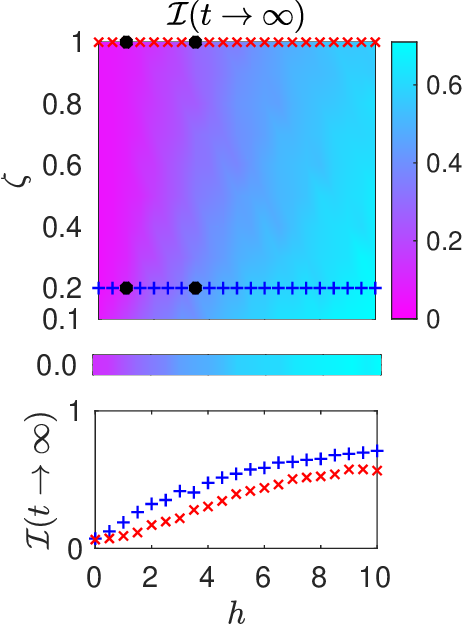}
\includegraphics[width=4.2cm,valign=M]{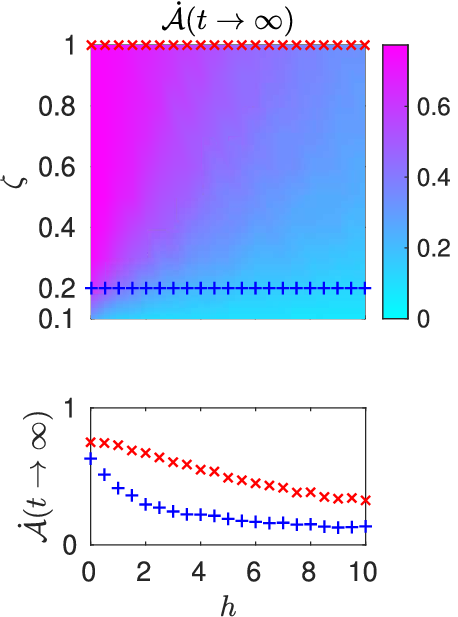}
\caption{(Left panel) Steady-state imbalance, $\mathcal{I}(t\to\infty)$ defined in Eq.~(\ref{eq:I_main}), as a function of disorder strength $h$ and quantum-jump-fugacity $\zeta$. The lower strip corresponds to $\zeta=0$.
(Right panel) Steady-state rate of dynamical activity, $\dot{\mathcal{A}}(t\to\infty)$ defined in Eq.~(\ref{eq:A_main}).
The lower panels correspond to the cuts at $\zeta = 0.2$ and $\zeta=1$ indicated in the upper panels.
The dynamics are generated by Eq.~(\ref{eq:Lindblad_zeta_main}). For imbalance (resp. dynamical activity), we consider a system of $L=10$ (resp. $L=8$) sites averaged over $100$ (resp. 160) disorder samples.
The four black dots in the upper left panel correspond to the parameters at which the transient time-dynamics are produced in Fig.~\ref{fig:ED_timedynamics}.
}
\label{fig:ED_steadystate}
\end{figure}

\paragraph*{Dynamical signatures.}
Strong disorder slows down the dynamics by raising the energetic barriers that suppress the inter-site hopping. A hallmark of localized dynamics is the ever-lasting memory of their initial conditions.
We choose to work with the charge density wave initial state $\rho(0)=|1,0,\cdots,1,0\rangle\langle 1,0,\cdots,1,0|$ which is a product state and a steady-state of the $\zeta$-deformed gain-loss dynamics in the absence of particle hopping, $J=0$.
We numerically integrate the subsequent dynamics generated by Eq.~(\ref{eq:Lindblad_zeta_main}) by employing a standard fourth-order Runge-Kutta algorithm (RK45). We quantify the fate of the staggered order present in the initial state $\rho(0)$ by computing the dynamics of the so-called imbalance~\cite{Schreiber2015Observation,Sierant2022Challenges}
\begin{align} \label{eq:I_main}
 \mathcal{I}(t)=\frac{\sum_{i=1}^{L} (-1)^{i+1} \mathrm{Tr}\left[b_i^\dagger b_i \rho_\zeta(t)\right]}{\sum_{i=1}^{L}\mathrm{Tr}\left[b_i^\dagger b_i \rho_\zeta(t)\right]}\,.
\end{align}
This is a directly observable quantity, $-1 \leq \mathcal{I}(t) \leq 1$, with $\mathcal{I}(t=0) = 1$.
Additionally, exploiting the formal analogy between full-counting statistics (FCS) in grand-canonical ensembles~\cite{Garrahan2010Thermodynamics} and the quantum trajectories ensemble interpretation of Lindblad dynamics~\cite{Manzano2022Quantum,Landi2023Current}, we monitor the rate of dynamical activity~\cite{Garrahan2011Quantum,Landi2023Current}
\begin{align} \label{eq:A_main}
\dot{\mathcal{A}}(t) = \frac{1}{\zeta} \partial_t \langle n(t) \rangle_\zeta\,,
\end{align}
where $ \langle n(t) \rangle_\zeta$ is the number of quantum jumps occurring between time $t=0$ to $t$  averaged over the quantum trajectories generated by $\mathcal{L}_\zeta$.
For the standard Lindblad evolution $\zeta=1$, the steady-state rate of dynamical activity is directly related to the imbalance as $\dot{\mathcal{A}}(t\to\infty) =  \gamma \, L \left[1 - \mathcal{I}(t\to\infty)\right]$. For $\zeta < 1$, $\dot{\mathcal{A}}(t\to\infty)$ involves additional contributions from two-time jump correlations. Details of this connection to FCS are provided in the Supplementary Material~[\onlinecite{SM}].
 
In Fig.~\ref{fig:ED_steadystate}, we present both the steady-state imbalance $\mathcal{I}(t\to\infty)$ and the rate of dynamical activity $\dot{\mathcal{A}}(t\to\infty)$ as a function of the disorder strength $h$ and the quantum-jump fugacity $\zeta$.
The results are consistent with those obtained from spectral statistics. For the standard Lindblad evolution $\zeta = 1$, we find a clear transition from a steady state with vanishing imbalance and a finite rate of dynamical activity at weak disorder to a steady state with imbalance close to unity and a vanishing rate of activity.
When $\zeta$ is reduced, the location of this transition is shifted to lower disorder strengths, confirming once again that quantum jumps tend to destabilize the localized regime.
Contrary to the results of the spectral statistics above, these dynamical indicators are not prone to subtleties involving symmetry sectors.
At finite but very small $\zeta$, the time it takes to reach the steady state diverges since the typical timescale between two jumps can be roughly estimated to be $\tau \sim 1/\gamma\zeta$. Indeed, the limits $t\to\infty$ and $\zeta \to 0$ are generically not expected to commute.
This results in significant numerical challenges in capturing the steady state and we do not provide data in the regime $\zeta \ll 1$.
To better illustrate the influence of the disorder strength $h$ and the quantum-jump fugacity $\zeta$, the lower panels of Fig.~\ref{fig:ED_steadystate} show the steady-state imbalance and rate of dynamical activity as a function of $h$ for two representative values of $\zeta$.

\begin{figure}
\centering
\includegraphics[width=8.4cm]{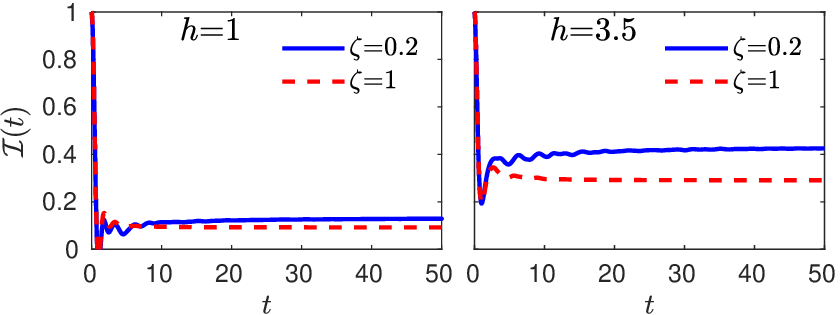}
\caption{Time dynamics of the imbalance $\mathcal{I}(t)$ defined in Eq.~(\ref{eq:I_main}) for representative values of the quantum-jump fugacity $\zeta$ and the disorder strength $h$ given in the legend.
The data are produced by numerically-exact integration of Eq.~(\ref{eq:Lindblad_zeta_main}) for a system of $L=10$ sites and averaged over $100$ disorder samples.
}
\label{fig:ED_timedynamics}
\end{figure}

\begin{figure}
\centering
\includegraphics[width=8.4cm]{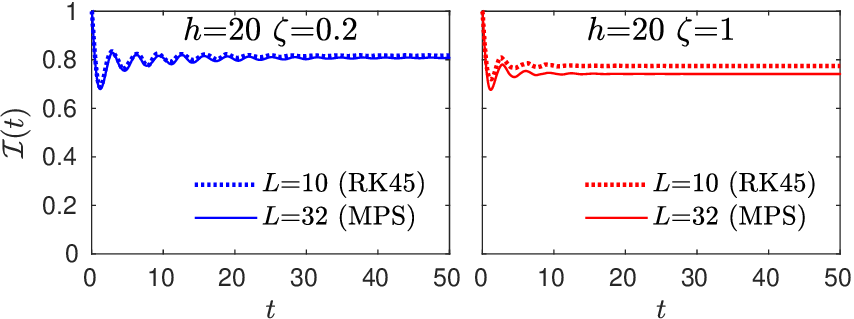}
\caption{Time dynamics of the imbalance $\mathcal{I}(t)$ for a very large system, $L=32$, deep in the localized regime, $h=20$, and for representative values of the  quantum-jump fugacity $\zeta$ given in the legend.
The data are produced by means of a time-dependent matrix product state (MPS) technique and averaged over $100$ disorder samples. The dashed lines correspond to the results obtained by numerically-exact integration of Eq.~(\ref{eq:Lindblad_zeta_main}) for a system of $L=10$ sites.
}
\label{fig:MPS_timedynamics}
\end{figure}

In Fig.~\ref{fig:ED_timedynamics}, we show the transient dynamics of imbalance from the initial state till the steady state for representative values of the disorder strength $h$ and the quantum-jump fugacity $\zeta$. 
The steady-state values increase with $h$ and decrease with $\zeta$. 
The timescale of the approach to the steady state is dictated by the inverse of the minimum Liouvillian gap~\footnote{The Liouvillian gap is defined as minus the real part of the difference of eigenvalues with largest and next largest real parts.}. We have used this spectral information to ensure the convergence of all steady-state results presented in this Letter. 
While the system sizes presented so far, up to $L = 10$, are state-of-the-art when it comes to exactly computing the dynamics of open quantum systems, they are still relatively small owing to the challenges posed by numerical time integration.
To firmly assert the influence of quantum jumps on the localized regime, we resort to a time-dependent matrix product state (MPS) technique that allows us to reach much larger system sizes, up to $L = 32$.
In practice, we implemented a  time-evolving block decimation (TEBD) of a matrix product density operator (MPDO) representation of the $\zeta$-deformed Lindblad evolution in Eq.~(\ref{eq:Lindblad_zeta_main}). See the Supplementary Material~[\onlinecite{SM}] for details.
The results are averaged over $100$ disorder samples.
This technique produces reliable results deep in the localized regime and we work at $h=20$ where convergence is achieved with a maximal bond dimension of $\chi = 2^7$. In Fig.~\ref{fig:MPS_timedynamics}, we show the transient dynamics of the imbalance from the initial state till the steady state for representative values of the quantum-jump fugacity $\zeta$. The MPS results entirely validate the previous results obtained by numerically-exact integration of systems of smaller sizes.

\paragraph*{Conclusion and discussion.}
We started from a disordered many-body system that already exhibited a localized regime and found that postselection protocols can facilitate localization at lower disorder strengths.
This is different from the measurement-induced phase transitions (MIPT)~\cite{Fisher2019MIPT,Huse2020purification,Pixley2020MIPT} where
repeated measurements can localize featureless systems such as random unitary circuits~\cite{Skinner2019Measurement} or free fermions~\cite{Buchhold2021Effective}
but are facing a major experimental challenge as they rely on generating and recording a large number of measurement trajectories. 
In our case, the $\zeta$-deformed Lindblad offers both a spectral and a dynamical window into the localized phase.
This approach not only harnesses standard methods of full counting statistics to the study of Lindblad dynamics, but it is also physically realizable by means of realistic postselection protocols in quantum optical setups.
It can easily be adapted to other systems of interest in condensed matter and quantum optics, such as open quantum spin chains, and driven-dissipative Jaynes-Cummings Hubbard systems, to name a few.

\paragraph*{Acknowledgments.}
We thank F. Minganti and F. Ferrari for useful discussions.
H.K.Y., C.A., and M.K. are grateful for the support from the Project No. 6004-1 of the Indo-French Centre for the Promotion of Advanced Research (IFCPAR). 
M.K. acknowledges support from the SERB Matrics Grant No. MTR/2019/001101 and the SERB VAJRA faculty scheme No. VJR/2019/000079. 
M.K. acknowledges support from the Department of Atomic Energy, Government of India, under Project No. RTI4001.
C.A. acknowledges the support from the French ANR ``MoMA'' Project No. ANR-19-CE30-0020.
M.K. is grateful for the hospitality of the  \'Ecole Normale Sup\'erieure (Paris) and the \'Ecole Polytechnique F\'ed\'erale de Lausanne.

\paragraph*{Author contributions.}
S.G. and H.K.Y. have equally contributed to producing the numerical data presented in this work.

\bibliography{references.bib}
\end{document}


\newcommand{\titlename}{{Supplementary material for}\\ ``Quantum jumps in driven-dissipative disordered many-body systems''}
\title{\titlename}
\author{Sparsh Gupta}
\email{sparsh.gupta@icts.res.in}
\affiliation{International Centre for Theoretical Sciences, Tata Institute of
Fundamental Research, 560089 Bangalore, India}
\author{Hari Kumar Yadalam}
 \email{hyadalam@uci.edu}
\affiliation{Laboratoire de Physique de l'\'Ecole Normale Sup\'erieure, ENS, Universit\'e PSL, CNRS, Sorbonne Universit\'e, Universit\'e Paris Cit\'e, F-75005 Paris,  France}
\affiliation{International Centre for Theoretical Sciences, Tata Institute of
Fundamental Research, 560089 Bangalore, India}
\affiliation{Department of Chemistry, University of California, Irvine, CA 92614, USA}
\affiliation{Department of Physics and Astronomy, University of California, Irvine, CA 92614, USA}
\author{Manas Kulkarni}
 \email{manas.kulkarni@icts.res.in}
\affiliation{International Centre for Theoretical Sciences, Tata Institute of Fundamental Research, 560089 Bangalore, India}
\author{Camille Aron}
 \email{aron@ens.fr}
  \affiliation{Institute of Physics, \'Ecole Polytechnique F\'ed\'erale de Lausanne (EPFL), CH-1015 Lausanne, Switzerland}
\affiliation{Laboratoire de Physique de l'\'Ecole Normale Sup\'erieure, ENS, Universit\'e PSL, CNRS, Sorbonne Universit\'e, Universit\'e Paris Cit\'e, F-75005 Paris,  France}
\date{\today}

\keywords{Quantum chaos, impurity model}

\maketitle

\tableofcontents 


\section{Construction of the $\zeta$-deformed theory}\label{app:QJT}
In this Section, we first briefly review the quantum jump trajectory interpretation of standard Lindblad master equations. Later, we discuss the deformation of the Lindblad equation introduced in the main manuscript in the language of quantum jump trajectories. Finally, we connect this deformation to physical post-selection protocols and non-Hermitian Hamiltonians.

\subsection{Quantum jump trajectory interpretation of standard Lindblad dynamics}
Let us consider a generic open quantum many-body system described by the following standard Lindblad master equation~\cite{breuer2002theory,carmichael2009open,daley2014quantum}
\begin{align}
\label{eq:Linblad_original}
\partial_t \rho(t)&= \mathcal{L}\rho(t)\,,
\end{align}
with the Liouvillian
\begin{align} \label{eq:Linbladian_original}
 \mathcal{L}\,\star&= -\rmi \left[H,\,\star\right]+\sum_{\alpha=1}^{M}\mathcal{D}[O_\alpha]\,\star\,,
\end{align}
where $H$ is the (Hermitian) Hamiltonian, $\mathcal{D}[O]\,\star := O\star O^\dagger -\frac{1}{2}\left\{O^\dagger O,\,\star\right\}$ is the standard Lindblad dissipator, and the $O_\alpha$'s are the jump operators where $\alpha=1, 2, \ldots M$ labels the dissipation channels.
Let us start from a generic initial state, described by a statistical mixture of pure states,
\begin{align} \label{eq:init}
\rho(0)=\sum_{m} p_{m} |\psi_{m} (0)\rangle \langle \psi_{m}(0)|\,, \mbox{ with }  \langle  \psi_{m} (0)| \psi_{m} (0) \rangle=1\,, \ p_m >0\,, \mbox{ and } \sum_m p_m =1\,.
\end{align}
At time $t$, the state of the system is given by
\begin{align} \label{eq:Linblad_evol}
 \rho(t)&= \rme^{\mathcal{L} t}\rho(0) \,.
\end{align}
The standard quantum-trajectory interpretation of the Lindblad dynamics~\cite{dalibard1992wave,molmer1993monte,dum1992monte,molmer1996monte,plenio1998the,carmichael2009open,daley2014quantum} involves separating the term responsible for the quantum jumps as
\begin{align}
 \mathcal{L} = \mathcal{L}_{0}+\mathcal{L}_{\rm J} \,,
 \end{align}
with the quantum-jump contribution reading
 \begin{align}
  \mathcal{L}_{\rm J}\, \star := \sum_{\alpha=1}^M O_{\alpha}\star O_{\alpha}^\dagger \,,
  \end{align}
and $ \mathcal{L}_{0}$ containing both the unitary dynamics generated by $H$ and the non-Hermitian contribution from the dissipators,  
\begin{align}
 \mathcal{L}_{0}\,\star :=-\rmi \tilde{H} \,\star+\star \, \rmi\tilde{H}^\dagger \,.
 \end{align}
Here, the effective non-Hermitian Hamiltonian $\tilde H$ is defined as
 \begin{align} \label{eq:H_tilde_2}
 \tilde{H} := H - \frac{\rmi}{2} \sum_{\alpha=1}^MO_{\alpha}^\dagger O_{\alpha} \,.
\end{align}
The time evolution of the density matrix given in Eq.~(\ref{eq:Linblad_evol}) can now be formally re-expressed as a
 \begin{align}
 \label{eq:rho_t2}
 \rho(t)&= \sum_{n=0}^{\infty} \int_{0}^{t} \rmd\tau_{n}\cdots\int_{0}^{\tau_2} \rmd\tau_1
\rme^{\mathcal{L}_{0}(t-\tau_n)}\mathcal{L}_{\rm J}
\rme^{\mathcal{L}_{0}(\tau_n-\tau_{n-1})}\cdots
\rme^{\mathcal{L}_{0}(\tau_2-\tau_1)}\mathcal{L}_{\rm J}
\rme^{\mathcal{L}_{0}\tau_1} \rho(0)\,,
 \end{align}
where we recall that
\begin{align}
\label{eq:exp_rel}
\rme^{\mathcal{L}_{0}\lambda}\, \star  =\rme^{-\rmi \tilde{H}\lambda} \,\star \, \rme^{\rmi\tilde{H }^\dagger \lambda} 
 \end{align}
for any parameter $\lambda$. The expression in Eq.~(\ref{eq:rho_t2}) can be interpreted as a Dyson series that sums over all the possible quantum jumps interrupting the dynamics generated by $\mathcal{L}_0$. The $n=0$ term corresponds to the non-Hermitian (no-jump) evolution.
Equation~(\ref{eq:rho_t2}), along with Eq.~(\ref{eq:init}), can be further rewritten as
 \begin{align} \label{eq:rho_t}
 \rho(t) = \sum_{n=0}^\infty \rho_n(t)\,,
 \end{align}
where the operator $\rho_n(t)$ is the contribution to the density matrix corresponding to evolution with a fixed number $n$ of quantum jumps occurring between time $t=0$ and $t$ and that reads
  \begin{align}
   \rho_0(t) &=  \rme^{\mathcal{L}_0 t} \rho(0) \,, \\
 \rho_n(t ) &= \sum_{m}p_m \hspace{-1.5em} \underbrace{\sum_{\alpha_1=1}^{M}\cdots\sum_{\alpha_n=1}^M \int_{0}^{t}
 \rmd\tau_{n}\cdots\int_{0}^{\tau_2} \rmd\tau_1 }_{\mbox{sum over all trajectories with $n$ jumps}}  \hspace{-1.5em} 
  P_m^{\, \boldsymbol{\alpha}_n,\, \boldsymbol{\tau}_n}(t)
  \big{|} \psi_m^{\, \boldsymbol{\alpha}_n,\, \boldsymbol{\tau}_n}(t) \big{\rangle}
  \big{\langle} \psi_m^{\, \boldsymbol{\alpha}_n,\, \boldsymbol{\tau}_n}(t) \big{|}\,, \ n \geq 1\,.  \label{eq:rho_n}
\end{align}
The symbols $\boldsymbol{\alpha}_n$ and $\boldsymbol{\tau}_n$ stand for the sequence of jump channels $(\alpha_1, \alpha_2, \ldots, \alpha_n)$ and jump times $(\tau_1, \tau_2, \ldots, \tau_n)$, respectively, and we have used Eq.~(\ref{eq:exp_rel}) in writing Eq.~(\ref{eq:rho_n}). Along with the initial state index $m$, these two sequences define a single quantum trajectory between time $t=0$ and $t$ with a total of $n$ quantum jumps.
$ \big{|} \psi_m^{\, \boldsymbol{\alpha}_n,\, \boldsymbol{\tau}_n}(t) \big{\rangle} $ is the normalized conditional wave function of the system at time $t$ which started in state $|\psi_m(0)\rangle$ and underwent an evolution with the precise sequence of jump channels $\boldsymbol{\alpha}_n$ and jump times $\boldsymbol{\tau}_n$. It reads
\begin{align} \label{eq:psi}
\big{|} \psi_m^{\, \boldsymbol{\alpha}_n,\, \boldsymbol{\tau}_n}(t) \big{\rangle} = \frac{\big{|} \tilde{\psi}_m^{\, \boldsymbol{\alpha}_n,\, \boldsymbol{\tau}_n}(t) \big{\rangle}}{\sqrt{P_m^{\, \boldsymbol{\alpha}_n,\, \boldsymbol{\tau}_n}(t)}}\,,
\end{align}
where
\begin{align}
\big{|} \tilde \psi_m^{\, \boldsymbol{\alpha}_n,\, \boldsymbol{\tau}_n}(t) \big{\rangle} := \rme^{-\rmi \tilde{H}(t-\tau_n)}O_{\alpha_{n}}\rme^{-\rmi \tilde{H}(\tau_n-\tau_{n-1})}O_{\alpha_{n-1}}\cdots \rme^{-\rmi \tilde{H}(\tau_2-\tau_1)}O_{\alpha_1}\rme^{-\rmi \tilde{H}\tau_1}|\psi_m(0)\rangle\,,
\end{align}
and $P_m^{\, \boldsymbol{\alpha}_n,\, \boldsymbol{\tau}_n}(t)$ is the probability of a quantum trajectory introduced above, given by
\begin{align} \label{eq:P_t}
P_m^{\, \boldsymbol{\alpha}_n,\, \boldsymbol{\tau}_n}(t)&=\big{\langle} \tilde \psi_m^{\, \boldsymbol{\alpha}_n,\, \boldsymbol{\tau}_n}(t)  \big{|} \tilde \psi_m^{\, \boldsymbol{\alpha}_n,\, \boldsymbol{\tau}_n}(t) \big{\rangle}.
\end{align}
Importantly, the fact that $\rho_n(t)$ can be written in the form of $\sum_\mu P_\mu | \psi_\mu \rangle \langle \psi_\mu |$ with $P_\mu \geq 0$ ensures its positive semi-definiteness.
$\mathrm{Tr} \, \rho_n(t)$ is precisely the probability of having experienced $n$ quantum jumps from time $t=0$ to $t$. This gives a probabilistic meaning to the quantum trajectories.
The special case of trajectories with no jump ($n=0$) corresponds to the evolution  generated by the non-hermitian Hamiltonian $\tilde H$ in Eq.~(\ref{eq:H_tilde_2})~\cite{Konstantin2014Comparison,Hamazaki2019Non,mcdonald2022nonequilibrium,Yamamoto2022Universal}:
\begin{align}
| \psi(t)\rangle = \frac{\rme^{-\rmi\tilde{H}t}|\psi(0)\rangle}{\sqrt{\langle  \psi(t) | \psi(t)\rangle }}\,.
\end{align}
Notably, taking the time-derivative of Eq.~(\ref{eq:rho_n}), one may check that the dynamics of the operators $\rho_n(t)$ follow
\begin{align} \label{eq:rho_n_dyn}
\partial_t \rho_n(t) &= \mathcal{L}_0 \rho_n(t) + \mathcal{L}_{\rm J}  \rho_{n-1}(t) \mbox{   for } n\geq 1\,, \\
\partial_t \rho_0(t) &= \mathcal{L}_0 \rho_0(t)\,.
\end{align}
This expresses the fact that a quantum state at time $t$ that is the result of $n-1$ jumps may either evolve linearly to time $t+\rmd t$  under the action of $\tilde H$ or be subject to a $n^{\rm th}$ jump.

\subsection{$\zeta$-deformed Lindblad dynamics}
Starting from Eq.~(\ref{eq:rho_t}), and its quantum trajectory interpretation, we now generalize the trajectory ensemble to a grand canonical ensemble by introducing a quantum jump fugacity $\zeta$, with $0 \leq \zeta \leq 1$,  which weights the trajectories with different numbers of jumps $n$.
From this perspective, the number of quantum jumps $n$ is analogous to the number of particles in the standard construction of statistical mechanics. The corresponding grand-canonical density matrix is constructed as
 \begin{align}
\rho_\zeta(t):=  \frac{1}{Z_\zeta(t)} \sum_{n=0}^{\infty} \zeta^n \rho_n(t) \,, \mbox{  where  }Z_\zeta(t) := \sum_{n=0}^{\infty} \zeta^n \mathrm{Tr} [\rho_n(t)]  \label{eq:rho_zeta}
\end{align}
is the grand-canonical partition function, ensuring $\mathrm{Tr}[\rho_\zeta(t) ]  = 1$ at all times and where $\rho_n(t)$ is given in Eq.~(\ref{eq:rho_n}).
The original density matrix $\rho(t)$ in Eq.~(\ref{eq:rho_t}) is recovered by setting $\zeta=1$ in Eq.~(\ref{eq:rho_zeta}). The no-jump case is recovered by setting $\zeta=0$.

The time evolution of the grand-canonical density matrix $\rho_\zeta(t)$ introduced in Eq.~(\ref{eq:rho_zeta}) can be computed by using the conditional evolution Eq.~(\ref{eq:rho_n_dyn}). One obtains the following $\zeta$-deformed Lindblad master equation
\begin{align}
\label{eq:Lindblad_zeta}
\partial_t  \rho_{\zeta}(t)&= \Big{(}
 \mathcal{L}_{\zeta}-\mathrm{Tr}\left[\mathcal{L}_{\zeta}\rho_{\zeta}(t)\right]\Big{)} \rho_{\zeta}(t),
\end{align}
where we introduced the $\zeta$-deformed Liouvillian
\begin{align}\label{eq:Liouvillian_zeta}
 \mathcal{L}_{\zeta} := \mathcal{L}_{0} + \zeta \mathcal{L}_{\rm J}
= -\rmi\left[H,\star\right]+\sum_{\alpha}^{}\left[\zeta O_\alpha\star O_\alpha^\dagger -  \frac{1}{2}\left\{O_\alpha^\dagger O_\alpha,\star\right\}\right].
\end{align}
The evolution in Eq.~(\ref{eq:Lindblad_zeta}) interpolates between the no-jump evolution at $\zeta=0$ and the original Lindblad master equation in Eq.~(\ref{eq:Linblad_original}) at $\zeta=1$.
Note that the trace term in Eq.~(\ref{eq:Lindblad_zeta}) stems from the factor $Z_\zeta(t)$ in Eq.~(\ref{eq:rho_zeta}) and we used $\partial_t Z_\zeta(t) = Z_\zeta(t) \, \mathrm{Tr} [\mathcal{L}_\zeta \rho_\zeta(t)]$. This trace term ensures the trace-preserving property of the dynamics of $\rho_\zeta(t)$ and makes the $\zeta$-deformed Lindblad master equation non-linear.
However, the density matrix in the $\zeta$-deformed theory can be linearly related to the initial density matrix of the non-deformed theory via
\begin{align} \label{eq:rho_zeta_2}
 \rho_\zeta(t) = \frac{1}{Z_\zeta(t)} \rme^{\mathcal{L}_\zeta t} \rho(0)\,.
\end{align}
Equation~(\ref{eq:rho_zeta_2}) may be checked by direct substitution into Eq.~(\ref{eq:Lindblad_zeta}) and implies a re-expression of the $\zeta$-deformed partition function as 
\begin{align} \label{eq:Z_zeta_2}
Z_\zeta(t) = \mathrm{Tr}[ \rme^{\mathcal{L}_\zeta t} \rho(0) ]\,.
\end{align}
Moreover, Eq.~(\ref{eq:rho_zeta_2}) shows that the operator $\rho_\zeta(t)$ inherits its Hermiticity property from $\rho(0)$. The positive semi-definite property of $\rho_\zeta(t)$ is ensured by the probabilistic interpretation of $\rho_n(t)$ in terms of quantum trajectories discussed in Eq.~(\ref{eq:rho_n}). Alternatively, this can be seen by recasting the infinitesimal evolution operator $1 - \mathcal{L} \delta t$ in a Krauss form and using the Krauss theorem.
Therefore, $\rho_\zeta(t)$ has all the expected properties of a well-defined density matrix.
Importantly, when evolving with the $\zeta$-deformed Liouvillian, the non-negative quantity $ \zeta^n \, \mathrm{Tr} \, \rho_n(t)$ is precisely the probability of having experienced $n$ quantum jumps from time $t=0$ to $t$. This interpretation is equivalent to the imperfect detection scheme that was discussed in Ref.~\cite{Minganti2020Hybrid,Ferri2023Entropy}.

\subsection{Post-selection interpretation at $\zeta < 1$}
Having argued the mathematical consistency of the $\zeta$-deformed theory in the above subsection, we now discuss how it can also be physically motivated. We have already presented one possible post-selection protocol to effectively realize the dynamics in the $\zeta$-deformed theory based on imperfect detectors. See Fig.~1 of the main manuscript. Here, we provide another possible implementation that was recently proposed in Ref.~\cite{Minganti2020Hybrid}.

Let us consider the original system, evolving with the original Liouvillian $\mathcal{L}$. Each dissipation channel can always be thought of as coupling the system to baths. We consider each channel coupled to two identical baths with coupling strength $\sqrt{\zeta}$ and $\sqrt{1-\zeta}$, respectively. This amounts in formally re-writing the original Liouvillian as
\begin{align} \label{eq:L_with_2_baths}
\mathcal{L} \, \star = - \rmi [H,\, \star] + \underbrace{\zeta \sum_{\alpha=1}^M \mathcal{D}[O_\alpha] \,\star }_{\mbox{non-monitored}} +  \underbrace{(1-\zeta)\sum_{\alpha=1}^M \mathcal{D}[O_\alpha]  \,\star}_{\mbox{monitored}} \,.
\end{align}
Instead of system-bath couplings, such a decomposition can also be seen as coupling each of the dissipation channels to a beam splitter which sends the particle into two detectors with probability $\zeta$ and $1-\zeta$, respectively.
The post-selection protocol consists in constantly monitoring the later bath/detector only. The results of this monitoring are used to post-select the quantum trajectories with no-jump with respect to the monitored bath/detector. The resulting dynamics is equivalent to erasing the jump term from the second Lindblad term in Eq.~(\ref{eq:L_with_2_baths}). It is now described  by
\begin{align}
 \mathcal{L}_{\zeta} = -\rmi\left[H,\star\right]+\sum_{\alpha}^{}\left[\zeta O_\alpha\star O_\alpha^\dagger -  \frac{1}{2}\left\{O_\alpha^\dagger O_\alpha,\star\right\}\right].
\end{align}
After normalizing observables with respect to the number of post-selected trajectories, in the limit of a large number of these, the results will match the ones computed from the dynamics governed by the trace-preserving Eq.~(\ref{eq:Lindblad_zeta}).

\section{Connection to full counting statistics} \label{eq:FCS}
In this Section, we exploit the connection of the $\zeta$-deformed theory to full counting statistics in order to relate the imbalance to an analog of a thermodynamic quantity, namely the activity. 

In the limit $\zeta=1$, the partition function introduced in Eq. (\ref{eq:rho_zeta}) can be seen as a moment generating function for the number of quantum jumps, or the activity in the language of full counting statistics, with $\zeta$ serving as the counting field. The time-dependent free energy, or the cumulant generating function, is defined as 
\begin{align}
\label{eq:fenergy}
F_{\zeta}(t):=\ln Z_{\zeta}(t) = \ln \left( \sum_{n=0}^{\infty} \zeta^n \, \mathrm{Tr} [\rho_n(t)] \right)\,.
\end{align}
Consequently, we get
\begin{align}
\label{eq:avqjumps}
\frac{\partial}{\partial \ln\zeta}F_{\zeta}(t)= \langle n(t) \rangle_{\zeta} =: \zeta \, \mathcal{A}(t)\,,
\end{align}
where $\langle n(t) \rangle_{\zeta}$ is the average number of quantum jumps in the $\zeta$-deformed theory,
\begin{align}
\langle n(t) \rangle_{\zeta} = \sum_{n=0}^{\infty} n\,  \zeta^n \, \mathrm{Tr} [\rho_n(t)] \geq 0
\end{align}
and $ \mathcal{A}(t)$ is the so-called dynamical activity.
Taking another derivative with respect to $\ln \zeta$ yields
\begin{align}
\frac{\partial}{\partial \ln\zeta}\langle n(t) \rangle_{\zeta} = \langle n(t)^2 \rangle_{\zeta}- \langle n(t) \rangle_{\zeta}^2 \geq 0\,,
\end{align}
implying that the average number of quantum jumps between time $t=0$ and $t$ is a non-decreasing function of $\zeta$.
Naturally, the number of quantum jumps is also a non-decreasing function of $t$. The associated rate of quantum jumps is directly related to the rate of dynamical activity, given by
\begin{align}
 \dot{\mathcal{A}}(t)  = \frac{1}{\zeta}\, \partial_t  \langle n(t) \rangle_{\zeta} \geq 0\,.
\end{align}
$\dot{\mathcal{A}}(t)$ can be computed by first taking the derivative of $F_{\zeta}(t)$ with respect to $\zeta$ and subsequently taking another derivative with respect to $t$
\begin{align}
\dot{\mathcal{A}}(t) =  \frac{\partial^2}{\partial t\,\partial \zeta}F_{\zeta}(t)\,.
\end{align}
 The former yields
\begin{align}
\label{eq:dzfenergy}
 \frac{\partial}{\partial \zeta}F_{\zeta}(t)=\frac{1}{Z_{\zeta}(t)} \int\limits_{0}^{t}\rmd\tau\ {\rm Tr}\left[ \rme^{\mathcal{L}_{\zeta}(t-\tau)}\mathcal{L}_{\rm J} \rme^{\mathcal{L}_{\zeta}\tau}\rho(0)\right]\,,
\end{align}
and the latter yields
\begin{align}
\dot{\mathcal{A}}(t) &={\rm Tr}\left[\mathcal{L}_{\rm J}\rho_\zeta(t)\right]
 -\frac{1-\zeta}{Z_{\zeta}(t)}
 \left( {\int\limits_{0}^{t}\rmd\tau\ {\rm Tr}\left[\mathcal{L}_{\rm J} \rme^{\mathcal{L}_{\zeta}(t-\tau)}\mathcal{L}_{\rm J} \rme^{\mathcal{L}_{\zeta}\tau}\rho(0)\right]} - {\rm Tr}\left[\mathcal{L}_{\rm J}\rho_\zeta(t)\right]\int\limits_{0}^{t}\rmd\tau\ {\rm Tr}\left[ \rme^{\mathcal{L}_{\zeta}(t-\tau)}\mathcal{L}_{\rm J} \rme^{\mathcal{L}_{\zeta}\tau}\rho(0)\right] \right) \nonumber \\
  &={\rm Tr}\left[\mathcal{L}_{\rm J}\rho_\zeta(t)\right]
 -\frac{1-\zeta}{Z_{\zeta}(t)} \int\limits_{0}^{t}\rmd\tau\ {\rm Tr}\left[\left(\mathcal{L}_{\rm J}-{\rm Tr}\left[\mathcal{L}_{\rm J}\rho_\zeta(t)\right]\right) \rme^{\mathcal{L}_{\zeta}(t-\tau)}\left(\mathcal{L}_{\rm J}-{\rm Tr}\left[\mathcal{L}_{\rm J}\rho_\zeta(\tau)\right]\right) \rme^{\mathcal{L}_{\zeta}\tau}\rho(0)\right]\,.\label{eq:dtdzfenergy}
\end{align}
This equation can finally be rewritten as,
\begin{align} \label{eq:dyn_act}
\dot{\mathcal{A}}(t)&= \big{\langle} \mathcal{L}_{\rm J}(t) \big{\rangle}-(1-\zeta) \int\limits_{0}^{t}\rmd\tau\ \big{\langle} \left[\mathcal{L}_{\rm J}(t)-\langle\mathcal{L}_{\rm J}(t)\rangle\right] \left[\mathcal{L}_{\rm J}(\tau)-\langle\mathcal{L}_{\rm J}(\tau)\rangle\right] \big{\rangle},
\end{align}
where in Eq.~(\ref{eq:dyn_act}) we introduced the notation 
\begin{equation}
\langle \cdots \rangle := \frac{1}{Z_\zeta(t)}\mathrm{Tr}\left[ \mathcal{T} \, \rme^{\int_{0}^{t}\rmd\tau \, \mathcal{L}_\zeta} \cdots \, \rho(0) \right] \, ,
\end{equation}
with the time-ordering operator $\mathcal{T}$ and $Z_\zeta(t)$ is given in Eq.~(\ref{eq:Z_zeta_2}).
The Eq.~(\ref{eq:dyn_act}) shows that for the standard Lindblad evolutions ($\zeta = 1$), the rate of dynamical activity can be simply expressed in terms of a single-time observable. In the generic case $\zeta < 1$, the rate of dynamical activity is reduced by additional contributions which can be expressed as a two-time connected correlation function of the jump operators.

In our specific model, the term $\langle \mathcal{L}_{\rm J}(t) \rangle$ is related to imbalance via
\begin{align}
\langle \mathcal{L}_{\rm J}(t) \rangle &=\sum_{i=1}^L\langle O_i^\dagger(t) O_i(t)\rangle = 2\gamma\left[ \frac{L}{2} - \sum\limits_{i=1}^{L} (-1)^{i+1} \, \langle b_i^\dagger(t) b_i(t)\rangle \right] = 2 \gamma \left(\frac{L}{2}- \langle I(t)\rangle \right)\,,
\end{align}
where we used the hard-core boson anti-commutation relations $\{ b_n, b_n^\dagger \} = 1$ and $ \langle I(t) \rangle$ is the numerator of the imbalance introduced in the main manuscript,
\begin{align} \label{eq:app_imb}
 \mathcal{I}(t) := \frac{\langle I(t) \rangle }{\langle N(t) \rangle}\,,
 \end{align}
 with
 \begin{align}
\langle I(t) \rangle  := \sum_{i=1}^{L}  (-1)^{i+1} \mathrm{Tr}\left[b_i^\dagger b_i \rho_\zeta(t)\right] \mbox{ and the total number }
\langle N(t) \rangle  := \sum_{i=1}^{L}\mathrm{Tr}\left[b_i^\dagger b_i \rho_\zeta(t)\right] \,.
\end{align}
At $\zeta = 1$, the two-time quantum jump correlator does not participate and we simply obtain $ \dot{\mathcal{A}}(t)= 2 \gamma \left(L/2- \langle I(t)\rangle \right)$.
In the limit $t\to\infty$, we have $\langle N(t\to\infty) \rangle = L/2$ (see Sect.~\ref{app:EoM}), and the rate of dynamical activity is  directly related to the imbalance as
\begin{align}
\dot{\mathcal{A}}(t\to\infty) =  \gamma \, L \left(1 - \mathcal{I}(t\to\infty)\right)\,, \qquad \zeta = 1\,. \label{eq:S40}
\end{align}
If we further assume that the disorder is strong enough to be in the localized regime,  $\mathcal{I}(t\to\infty)\to 1$ (see Sect.~\ref{app:EoM}) and $\dot{\mathcal{A}}(t\to\infty)\to 0$. The numerical results presented in Fig.~4 of the main manuscript are in perfect agreement with the identity in Eq.~(\ref{eq:S40}).

\section{Symmetries of $\mathcal{L}_\zeta$ in the gain-loss model}
In this Secion, we discuss the symmetries of the $\zeta$-deformed Liouvillian $\mathcal{L}_\zeta$ defined in Eq.~(\ref{eq:Liouvillian_zeta}) with the Hamiltonian introduced in the main manuscript that reads
\begin{align} \label{eq:H_2}
 H&= \sum_{i=1}^{L} h_i b_i^\dagger b_i-J \sum_{i=1}^{L-1}\left[b_i^\dagger b_{i+1} + b_{i+1}^\dagger b_i\right]  + U \sum_{i=1}^{L-1} b_i^\dagger b_i b_{i+1}^\dagger b_{i+1} \,,
\end{align}
where the $b_i$'s are hard-core bosons and the jump operators
\begin{align}
O_i &=\begin{cases} \sqrt{2\gamma} \ b_i^\dagger & \text{if} \ i \ \text{is odd}\\
      \sqrt{2\gamma} \ b_i & \text{if} \ i \ \text{is even} \,.
     \end{cases}
\end{align}
The isolated Hamiltonian given in Eq.~(\ref{eq:H_2}) is $U(1)$-symmetric, corresponding to a conservation of the total particle number $N=\sum\limits_{i=1}^{L} b_i^\dagger b_i$, that is $[H,\, N] = 0$.
On the other hand, the $\zeta$-deformed Liouvillian in Eq.~(\ref{eq:Liouvillian_zeta}) has a weak-$U(1)$ symmetry, 
\begin{align} \label{eq:weak_sym}
[\mathcal{L}_\zeta, \, \mathcal{N}_{-}] = 0 \,, \mbox{ where } \mathcal{N}_{-}  \, \star := \left[ N,\, \star\right]\,.
\end{align}
The operator $\mathcal{N}_{-}$, when acting on a state $|n, \alpha\rangle  \langle m,\beta| $ where $n$, $m$ are the quantum numbers associated to the $U(1)$ symmetry and $\alpha$, $\beta$ account for all other quantum numbers defining the state, counts the number difference on the ket and bra side of the state, \textit{i.e.} $n-m$. The weak symmetry translates into conserving this number difference upon acting with $\mathcal{L}_\zeta$, and therefore along the dynamics generated by $\mathcal{L}_\zeta$.

Let us now explicitly show the weak symmetry in Eq.~(\ref{eq:weak_sym}) using a generic state $\rho$. For the sake of simplicity, we only consider a single dissipative channel with jump operator $O = \sqrt{2\gamma} \, b^\dagger$, and the general case follows immediately.
We expand the commutators 
\begin{align}
[\mathcal{L}_\zeta, \, \mathcal{N}_{-}] \rho &= \mathcal{L}_\zeta (N \rho - \rho N) - \left[ N \mathcal{L}_\zeta(\rho) -  \mathcal{L}_\zeta(\rho) N\right] \\
& = \mathcal{L}_0 (N \rho - \rho N) - \left[ N \mathcal{L}_0(\rho) -  \mathcal{L}_0(\rho) N\right]
+ \zeta  \left\{ \mathcal{L}_{\rm J} (N \rho - \rho N) - \left[ N \mathcal{L}_{\rm J}(\rho) -  \mathcal{L}_{\rm J}(\rho) N\right] \right\} \nonumber \\
& = - \frac12 \left\{ O^\dagger O (N \rho - \rho N) + O^\dagger O (N \rho - \rho N) \right\} 
+ \zeta  \left\{ \mathcal{L}_{\rm J} (N \rho - \rho N) - \left[ N \mathcal{L}_{\rm J}(\rho) -  \mathcal{L}_{\rm J}(\rho) N\right] \right\} \nonumber \\
& = \zeta  \left\{ \mathcal{L}_{\rm J} (N \rho - \rho N) - \left[ N \mathcal{L}_{\rm J}(\rho) -  \mathcal{L}_{\rm J}(\rho) N\right] \right\} \nonumber \\
& = \zeta  \left\{ O (N \rho - \rho N) O^\dagger - \left[ N O \rho \, O^\dagger -  O \rho \, O^\dagger N\right] \right\} \nonumber \\
& = 0 \,, \nonumber
\end{align}
where we used $[H,N] = 0$ in the third line, and in the fourth line we used the fact that our choice of jump operators obeys $[O^\dagger O,N]$. In the sixth line, we used the commutation relations $[O,N] = - O$ and $[O^\dagger,N] =  O^\dagger$.

For the special case $\zeta=0$, $\mathcal{L}_{\zeta=0} = \mathcal{L}_0$ has an additional weak $U(1)$ symmetry 
\begin{align}
 \label{eq:weak_sym2}
[\mathcal{L}_{0}, \, \mathcal{N}_{+}] = 0 \,, \mbox{ where } \mathcal{N}_{+}  \, \star := \left\{ N,\, \star\right\}\,.
\end{align}
Indeed,
\begin{align}
[\mathcal{L}_0, \, \mathcal{N}_{+}] \rho &= \mathcal{L}_0 (N \rho + \rho N) - \left[ N \mathcal{L}_0(\rho) +  \mathcal{L}_0(\rho) N\right] \\
& = \mathcal{L}_0 (N \rho + \rho N) - \left[ N \mathcal{L}_0(\rho) +  \mathcal{L}_0(\rho) N\right]
\nonumber \\
& = - \frac12 \left\{ O^\dagger O (N \rho + \rho N) + (N \rho + \rho N)  O^\dagger O \right\} + \frac12
\left\{ N( O^\dagger O \rho + \rho  O^\dagger O  ) + ( O^\dagger O \rho + \rho  O^\dagger O  ) N \right\} 
 \nonumber \\
& = 0\,, \nonumber
\end{align}
where we also used $[H,N] = [O^\dagger O,N]= 0$.
For this $\zeta=0$ case, one can form linear combinations of the two weak $U(1)$ symmetry generators as $\mathcal{N}_{\mathrm{ket}/\mathrm{bra}}=\frac{\mathcal{N}_{+}\pm\mathcal{N}_{-}}{2}$, which count particle number on the ket and bra sides, respectively.

\section{Equation of motion for population and imbalance}\label{app:EoM}
In this Section, we present the equations of motion under the evolution generated by Eq.~(\ref{eq:Lindblad_zeta}) for the average total number of particles 
\begin{align}
\langle N(t) \rangle := \mathrm{Tr}\left[  \sum_{i=1}^N b_i^\dagger b_i \, \rho(t) \right]\,,
\end{align}
and
\begin{align}
\langle I(t) \rangle :=  \mathrm{Tr} \left[  \sum_{i=1}^N (-1)^{i+1} b_i^\dagger b_i \, \rho(t) \right]\,,
\end{align}
that are, respectively, the numerator and the denominator of the imbalance $\mathcal{I}(t)$ introduced in Eq.~(\ref{eq:app_imb}).
Defining the average bond current between sites $i$ and $i+1$ as 
\begin{align}
\mathcal{J}_{i}(t) :=-\rmi J \, \mathrm{Tr} \left[  ( b_{i}^{\dagger} b_{i+1}-b_{i+1}^{\dagger} b_{i}) \rho(t) \right]\,,
\end{align}
we obtain the set of equations
\begin{align}
\frac{\langle\dot{I}\rangle}{2 \gamma} &= (1-\zeta)\left(\langle {I}^{2}\rangle-\langle {I}\rangle^{2}\right)+ \zeta\left(\frac{L}{2}-\langle {I}\rangle\right) - \frac{1}{\gamma} \sum_{i=1}^{L-1} (-1)^{i+1}  \mathcal{J}_{i} \,, \label{eq:eom_N}\\
\frac{\langle\dot{N}\rangle}{2\gamma} &= (1-\zeta) \left(\langle  {N} {I}\rangle-\langle N\rangle\langle {I}\rangle \right) + \zeta \left(\frac{L}{2} - \langle N\rangle \right)  \label{eq:eom_I}\,.
\end{align}
Once the steady state is reached, the left-hand sides of these equations vanish. This implies, notably, that in the standard Lindblad evolution ($\zeta = 1$), the steady state is half-filled, \textit{i.e.} $\langle N(t\to\infty) \rangle = L/2$ and its imbalance $\mathcal{I}(t\to\infty) \to 1$ deep in the localized regime where one expects the staggered current in the right-hand side of Eq.~(\ref{eq:eom_N}) to vanish.
Since our initial state $|1,0, \ldots,1,0  \rangle$ is maximally imbalanced (with charge $L/2$) we have  $\mathcal{I}(t) = 1$ for all $t\geq 0$ deep in the localized regime.
On the other hand, in the non-Hermitian limit ($\zeta = 0$), Eq.~(\ref{eq:eom_I}) implies that the steady state is a pure Fock state of the form, \textit{e.g.}, $|1, 1, \ldots, 0, 1\rangle$.  
Deep in the localized regime where the staggered current is expected to vanish, the dynamics of $\langle I \rangle$ is completely arrested, \textit{i.e.} $\langle \dot I \rangle = 0$ for all $t \geq 0$, since the initial state $|1,0, \ldots,1,0  \rangle$ is an eigenstate of imbalance and of the non-Hermitian Hamiltonian. This implies that the imbalance $\mathcal{I}(t) = 1$ for all $t\geq 0$ deep in the localized regime in both the $\zeta = 1$ and
$\zeta=1$ regimes.

\section{Time-dependent matrix product state solver}
In this Section, we provide a pedagogical introduction to the matrix product density operators (MPDO) representation of states. Later, we detail our implementation of the time-evolving block decimation (TEBD) solver which we use to solve the dynamics of large systems, up to $L=32$. Lastly, we demonstrate the convergence of the results presented in Fig.~6 of the main manuscript.

\subsection{Matrix product density operators (MPDO) representation of states}

\begin{figure}
\centering
\includegraphics[width=.8\linewidth]{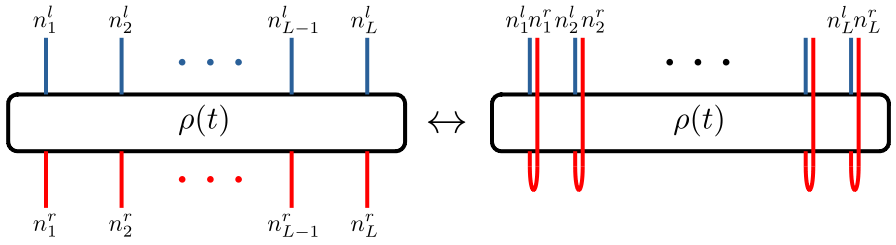}
\caption{ Two equivalent ways to represent the matrix elements of the density operator in the computational basis: (Left) ket and bra indices for all the sites are separately clubbed together, and (Right)  ket and bra indices for each site are clubbed together.
}
\label{fig:rho}
\end{figure}

The Hilbert space of a system of hard-core bosons on a one-dimensional lattice of $L$ sites is of dimension $2^L$. Consequently, the matrix representation of the density matrix is of at most $4^L$ real parameters. In the Fock state basis, this can be generically represented as
\begin{align}
 \rho &= \sum_{n_1^l,n_1^r=0}^{1}\cdots\sum_{n_L^l,n_L^r=0}^{1} \rho^{n_1^l,n_1^r;\cdots;n_L^l,n_L^r} |n_1^l,\cdots,n_L^l\rangle \langle n_1^r,\cdots,n_L^r|\,,
\end{align}
where $|n_1^l,\cdots,n_L^l\rangle$ are the Fock states spanning the Hilbert space. The superscripts $l$ and $r$ indicate ket and bra, respectively. $\rho^{n_1^l,n_1^r;\cdots;n_L^l,n_L^r}$ is a tensor with $2L$ indices $n_i^{l/r}$, $i = 1,\,\ldots,\, L$ taking values in $\{0,1\}$. $\rho^{n_1^l,n_1^r;\cdots;n_L^l,n_L^r}$ is represented on the right side of Fig.~\ref{fig:rho}.

\begin{figure}
\centering
\includegraphics[width=.8\linewidth]{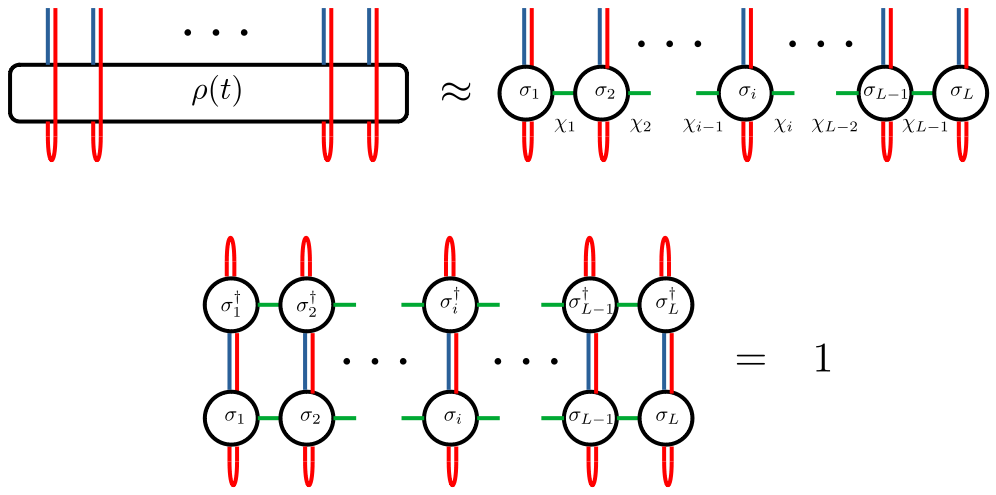}
\caption{ (Top) Matrix product density operator (MPDO) representation of the density matrix, and (Bottom) MPDO is conveniently normalized as ${\rm Tr}[\rho^\dagger \rho]=1$, rather than ${\rm Tr}[\rho]=1$.
}
\label{fig:mpdo}
\end{figure}

The Matrix Product Density Operator (MPDO) representation of such a density matrix is given by by~\cite{schollwock2011density,jaschke2018one,weimer2021simulation}
\begin{align}
 \rho^{n_1^l,n_1^r;\cdots;n_L^l,n_L^r} &= \sum_{\chi_1=1}^{D_1}\cdots \sum_{\chi_{L-1}=1}^{D_{L-1}} 
 {\sigma_1}_{\chi_0\chi_1}^{n_1^l,n_1^r}
 {\sigma_2}_{\chi_1 \chi_2}^{n_2^l,n_2^r}  
  \cdots
 {\sigma_{L-1}}_{\chi_{L-2}\chi_{L-1}}^{n_{L-1}^l,n_{L-1}^r}
 {\sigma_L}_{\chi_{L-1} \chi_L}^{n_L^l,n_L^r}\,.
\end{align}
This is depicted schematically in Fig.~\ref{fig:mpdo}. Here, at a given site $i$, the so-called site tensor ${\sigma_i}_{\chi_{i-1}\chi_i}^{n_i^l,n_i^r}$ is four-dimensional (three-dimensional at the boundaries. The subscript indices indicate the bond/auxiliary dimensions $\chi_{i} \in \{1,\cdots, D_i\}$ with $D_i\leq \min\{2^{2 i},2^{2(L-i)}\}$ for $i = 1,2, \, \ldots,\, L-1$ and $\chi_0 = \chi_L=1$, and quantify the maximal amount of operator entanglement entropy of the density operator~\cite{prosen2007operator}.
The denomination ``matrix product density operator'' comes from noting that, for a given value of the physical indices, $n_i^l, n_i^r$, each site tensor is a matrix and hence $ \rho^{n_1^l,n_1^r;\cdots;n_L^l,n_L^r}$ is given by a product of matrices, where the contraction is performed along the bond dimensions.

\begin{figure}
\centering
\includegraphics[width=.5\linewidth]{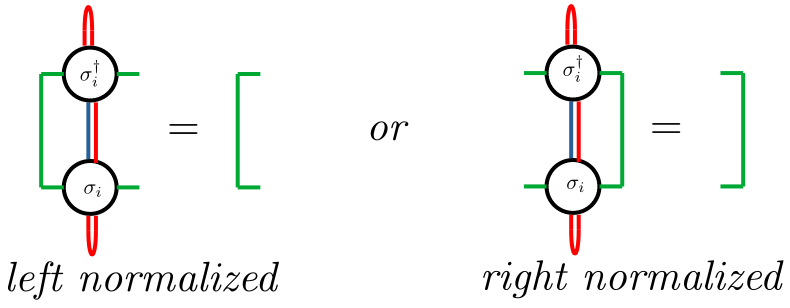}
\caption{ (Left) Site tensor is left normalized, and (Right) Site tensor is right normalized.
}
\label{fig:canonical}
\end{figure}

It is important to note that the matrix product density operator representation is not unique~\cite{schollwock2011density}. Indeed, let us consider $L-1$ invertible matrices $X_{i}$ of dimension $D_i \times D_i$ with $i = 1,2, \, \ldots,\, L-1$.
If we transform the site tensors with fixed physical dimension as,
\begin{align}
\left\{
\begin{array}{rl}
\sigma_1^{n_1^l,n_1^r} &\to \sigma_1^{n_1^l,n_1^r} X_1,  \\
\sigma_i^{n_i^l,n_i^r} &\to X_{i-1}^{-1}\sigma_i^{n_i^l,n_i^r} X_{i}  \ \mbox{  for }  i = 2,3 \, \ldots,\, L-1\,,\\
\sigma_L^{n_L^l,n_L^r} &\to X_{L-1}^{-1}\sigma_L^{n_L^l,n_L^r}\,,
\end{array}
\right.
\end{align}
the new site tensors give the same density matrix upon contraction. For various matrix product state algorithms, there are three typical representations which simplify the computations:
\begin{enumerate}
\item[(i)] \textit{Left canonical form} is such that all site tensors satisfy the orthonormality condition,
\begin{equation}
 \sum_{\chi_{i-1},n_i^l,n_i^r} {\sigma_i^*}_{\chi_{i-1} \chi_{i}}^{n_i^l,n_i^r}{\sigma_i}_{\chi_{i-1} \chi_{i}^\prime}^{n_i^l,n_i^r}=\delta_{\chi_i \chi_i^\prime}\,,
\end{equation}
depicted on the left side of Fig.~\ref{fig:canonical}.
\item[(ii)] \textit{Right canonical form} is such that all site tensors satisfy the orthonormality condition,
\begin{equation}
 \sum_{\chi_{i},n_i^l,n_i^r} {\sigma_i^*}_{\chi_{i-1} \chi_{i}}^{n_i^l,n_i^r}{\sigma_i}_{\chi_{i-1}^\prime \chi_{i}}^{n_i^l,n_i^r}=\delta_{\chi_{i-1} \chi_{i-1}^\prime}\,,
\end{equation}
depicted on the right side of Fig.~\ref{fig:canonical}.
\item[(iii)] \textit{Mixed canonical form} is such that for a specific site, say $i$, called the orthonormality center, all the preceding site tensors are in the left canonical form and all the following site tensors are in the right canonical form.
\end{enumerate}
Any MPDO can be brought into any of these canonical forms using QR decomposition~\cite{schollwock2011density}.
Notice that in the left and right canonical forms, the matrix product density operator representation of a density matrix is normalized to ${\rm Tr}\left[\rho^\dagger\rho\right]=1$. Once the steady-state or the time-dependent density matrix is obtained from the matrix product state algorithm, one divides the expectation value of observables by ${\rm Tr}\left[\rho\right]$.

The approximation methods for computing steady-state density matrices as well as time-propagating density matrices under generic Markovian evolutions using matrix product density operator representation involve approximating MPDOs by truncating them to a maximal bond dimension $\chi$, $\max \{D_1,\cdots,D_{L-1}\} \leq \chi$. With such an approximation, the density matrix is parameterized by at most $4 \times \chi^2 \times L$ (linear in system size) parameters as opposed to the exact representation which needs $4^L$ (exponential in system size) parameters.

\subsection{Dynamics of MPDO with time-evolving block decimation}

\begin{figure}
\centering
\includegraphics[width=1.\linewidth]{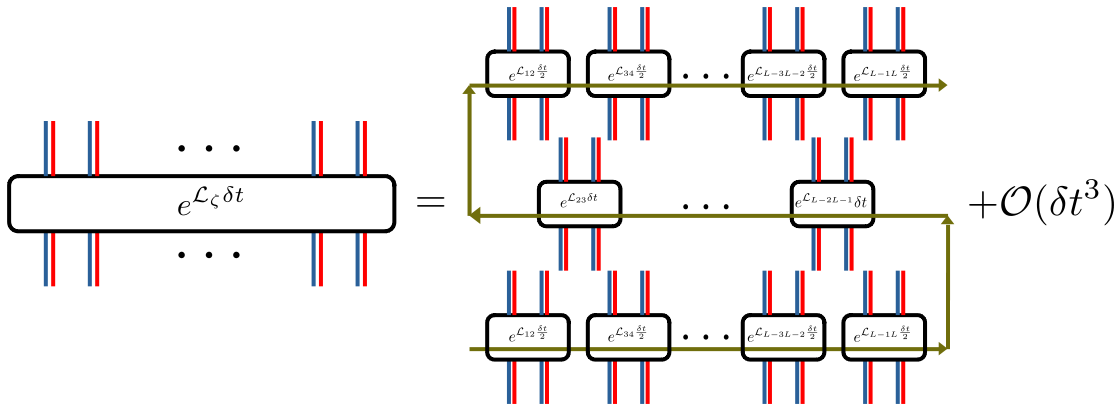}
\caption{Second-order approximation of the short-time propagator $  \rme^{\mathcal{L}_{\zeta}\delta\tau}$. Here, $L$ is assumed to be an even number.
}
\label{fig:trotter2}
\end{figure}

The time evolution of matrix product density operator representation of the density matrix, namely the site tensors $\sigma_i$'s, is performed using the generalization of the time-evolution by block decimation algorithm~\cite{vidal2003efficient,vidal2004efficient,schollwock2011density,paeckel2019time} proposed in Ref.~\cite{verstraete2004matrix}, also see the reviews in Refs.~\cite{jaschke2018one,weimer2021simulation}.
The time evolution by block decimation technique for time propagation involves Trotter approximating the short-time evolution operator. Here, we used second-order Trotterization~\cite{schollwock2011density,paeckel2019time}. For nearest neighbor Liouvillians such as the ones considered in this work, the Trotterization involves making use of the two-site structure of the Liouvillian, 
\begin{equation}
\mathcal{L}_{\zeta}=\sum_{i=1}^{L-1}\mathcal{L}_{i i+1}:=\mathcal{L}_{{\rm odd}}+\mathcal{L}_{{\rm even}}\, , 
\end{equation}
and approximating the short-time propagator as
\begin{align}
  \rme^{\mathcal{L}_{\zeta}\delta\tau}& =  \rme^{\mathcal{L}_{{\rm odd}}\frac{\delta\tau}{2}} \rme^{\mathcal{L}_{{\rm even}}\delta\tau} \rme^{\mathcal{L}_{{\rm odd}}\frac{\delta\tau}{2}} + \mathcal{O}\left((\delta \tau)^3\right)\,,
\end{align}
which is depicted in Fig.~\ref{fig:trotter2}.

The Trotter decomposition error is controlled by choosing the size of the time step, here $\delta\tau \leq 10^{-2}$. Each odd and even site propagator are direct product of two-site propagators,
\begin{align}
   \rme^{\mathcal{L}_{{\rm odd}}\tau}&= \otimes_{i \in {\rm odd}} \rme^{\mathcal{L}_{i i+1} \tau} \\
    \rme^{\mathcal{L}_{{\rm even}}\tau}&= \otimes_{i \in {\rm even}} \rme^{\mathcal{L}_{i i+1} \tau} \,.
\end{align}

\begin{figure}
\centering
\includegraphics[width=.5\linewidth]{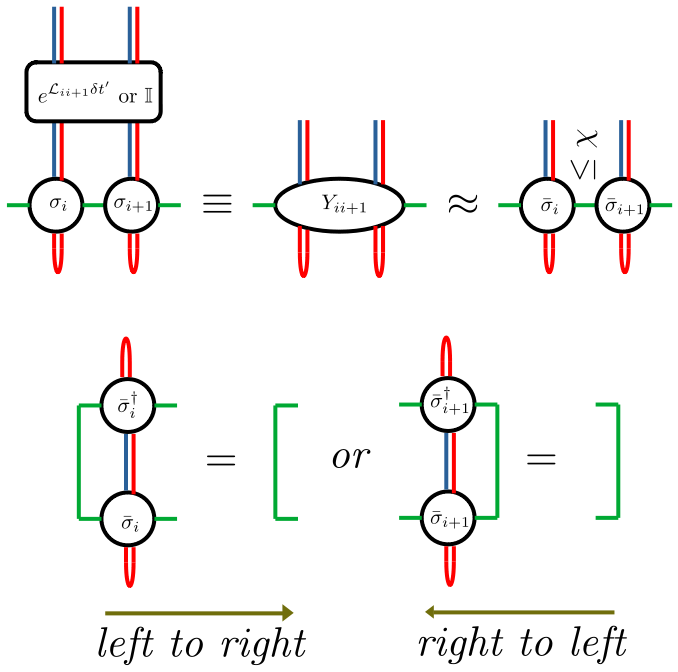}
\caption{(Top) Updating site tensor for a pair of neighboring sites, and (Bottom) Update is done such that the site tensor $\bar{\sigma}_i^{}$ ($\bar{\sigma}_{i+1}^{}$) is left (right)-normalized for left-to-right (right-to-left) sweep.
}
\label{fig:siteupdate}
\end{figure}

\begin{figure}
\centering
\includegraphics[width=.6\linewidth]{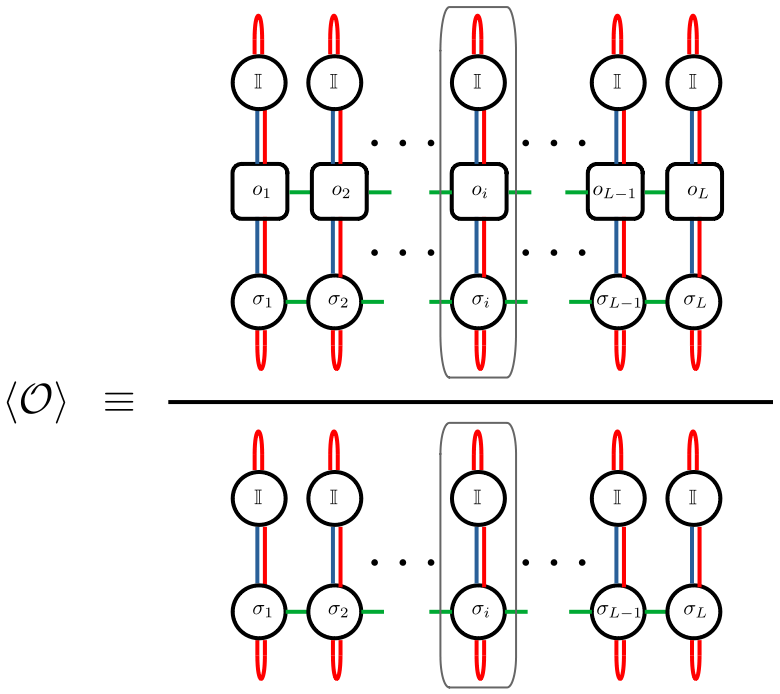}
\caption{ Expectations values of a generic superoperator expressed in matrix product superoperator form (MPSO) are computed by the above contraction.
}
\label{fig:expectation}
\end{figure}

This structure allows us the joint update of nearest-neighbor pairs of site tensors, say $\sigma_i$ and $\sigma_{i+1}$, by means of two-site gate evolution operators. In practice, we first construct the two-site tensor
\begin{align}
 {Y_{i,i+1}}_{\chi_{i-1}\chi_{i+1}}^{n_{i}^{l},n_{i}^{r};n_{i+1}^{l},n_{i+1}^{r}}= \sum_{n_{i}^{l\prime},n_{i}^{r\prime}} \sum_{n_{i+1}^{l\prime},n_{i+1}^{r\prime}}
[ \rme^{\mathcal{L}_{i i+1} \tau}]_{}^{n_{i}^{l},n_{i}^{r},n_{i+1}^{l},n_{i+1}^{r};n_{i}^{l\prime},n_{i}^{r\prime},n_{i+1}^{l\prime},n_{i+1}^{r\prime}} \sum_{\chi_{i}}  {\sigma_i}_{\chi_{i-1}\chi_{i}}^{n_{i}^{l\prime} n_{i}^{r\prime}}{\sigma_{i+1}}_{\chi_{i}\chi_{i+1}}^{n_{i+1}^{l\prime} n_{i+1}^{r\prime}}\,.
\end{align}
This is followed by re-expressing this two-site tensor as a contraction of the two updated single-site tensors, denoted by $\overline{\sigma}_i$ and $\overline{\sigma}_{i+1}$, as
\begin{align}
  & {Y_{i,i+1}}_{\chi_{i-1}\chi_{i+1}}^{n_{i}^{l},n_{i}^{r};n_{i+1}^{l},n_{i+1}^{r}}=\sum_{\overline{\chi}_{i}=1}^{\overline{D}_i}{{\overline\sigma}_i}_{\chi_{i-1}\overline{\chi}_{i}}^{n_{i}^{l} n_{i}^{r}}{\overline{\sigma}_{i+1}}_{\overline{\chi}_{i}\chi_{i+1}}^{n_{i+1}^{l} n_{i+1}^{r}}\,.
\end{align}
This can be performed using singular value decomposition~\cite{schollwock2011density,paeckel2019time}. This procedure is depicted in Fig.~\ref{fig:siteupdate}.  This gives back the matrix product representation of the density operator.
At this step, the updated bond dimensions of the tensors, $\overline{D}_{i}$, generically increases up to $16\,D_i$~\cite{schollwock2011density,paeckel2019time}.
To keep these growing bond dimensions under control, a truncation is performed to keep the growing bond dimensions bounded below by $\chi$, by choosing only singular values below a tolerance, here $10^{-16}$ not exceeding the total number of singular values retained to $\chi$, here $\chi=2^7$.
This way, the odd or even propagators can be applied to the matrix product density operator representation of the state by applying the constituent two-site gates.
If the truncation step is not performed, the two-site gates constituting the even or odd propagators can be applied in parallel.
However, as discussed above, to keep the growing bond dimensions under control, one has to perform a truncation step, which can be performed optimally only if the single site tensors left and right to the two sites under consideration are in the left and right canonical form, respectively.
This gives us a relatively simple procedure for propagating density matrix product states, often referred as the zip-up method~\cite{schollwock2011density,paeckel2019time}.
In this method, the odd site propagators are first applied to the matrix product density operator state one after the other, while keeping the state in the mixed canonical form after truncation and before proceeding to the next pair of sites.
This is performed for example from left to right of the chain.
This is then followed by a similar application of the even site propagators, from right to left.
Finally, the odd site propagator is applied from left to right. This sequence of steps is indicated by grey arrows in Fig.~\ref{fig:trotter2}. At the end of these steps, time is incremented by one time step.
The observable values can be computed at regular intervals of time~\cite{jaschke2018one,cui2015variational,casagrande2021analysis}. Expectation values of such observables expressed in matrix product superoperator form can be computed as shown in Fig.~\ref{fig:expectation}.

It must be noted that this time evolution by block decimation algorithm applied to matrix product density operator does not preserve the positive semi-definiteness of the density matrix. 
Therefore, one must choose $\delta t$ and $\chi$ sufficiently small and large, respectively, to make sure that the computed observable quantities are real. Here $\delta t \leq 10^{-2}$ and $\chi \geq 2^5$ was sufficient to keep the magnitude of the imaginary parts of physical observables below $10^{-5}$.

\begin{figure}
\centering
\includegraphics[width=15.6cm]{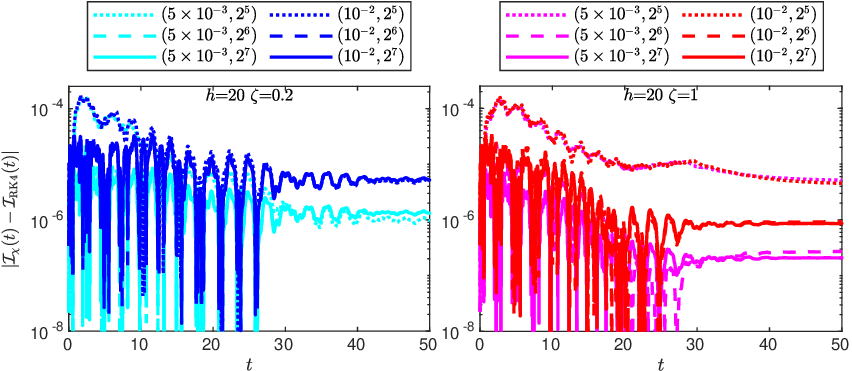}
\caption{Error in the imbalance as a function of time, $\mathcal{I}_{\chi}(t)-\mathcal{I}(t)$, computed using time-evolving block decimation (TEBD) with second-order Trotter decomposition compared with exact imbalance for a single realization of the disorder with strength $h=20$ and for $\zeta=0.2$ (left) and $\zeta=1$ (right) for a system of $L=8$ sites. The error is plotted for various values of the pair $(\delta t, \chi)$ given in the legend, which controls the accuracy of TEBD results.
\label{fig:MPS_benchmark1}
}
\end{figure}
\begin{figure}
\centering
\includegraphics[width=15.6cm]{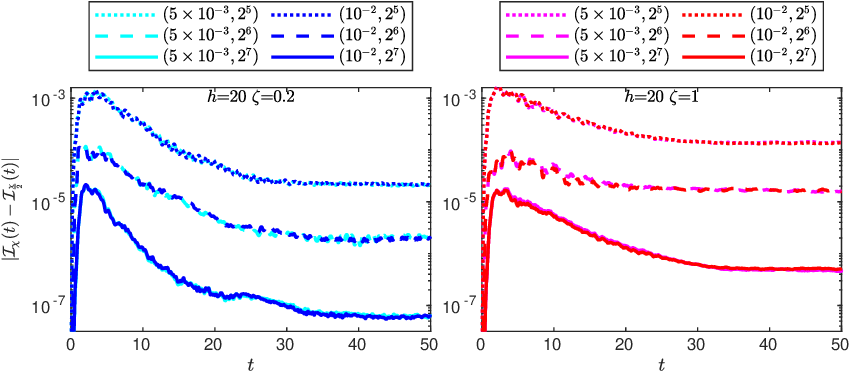}
\caption{Imbalance as a function of time, $\mathcal{I}_{\chi}(t)$, computed using time-evolving block decimation (TEBD) with second-order Trotter decomposition for a single realization of the disorder with strength $h=20$ and for $\zeta=0.2$ (left) and $\zeta=1$ (right) for system with $L=32$ sites. The imbalance is plotted for various values of the pair $(\delta t, \chi)$ given in the legend.
\label{fig:MPS_benchmark2}
}
\end{figure}

\subsection{Benchmark of time-evolving block decimation MPDO versus numerically-exact dynamics}
We employed the matrix product density operator (MPDO) representation and the time-evolving block decimation (TEBD) discussed in the previous subsections for computing the dynamics of imbalance in large-size systems ($L=32$), presented in Fig.~6 of the main manuscript.
The disorder averaged imbalance dynamics presented in Fig.~6 are obtained with a maximal bond dimension $\chi=2^6$ for $\zeta=0.2$, and $\chi=2^7$ for $\zeta=1$. The second-order Suzuki Trotter time step is chosen as $\delta t = \times 10^{-2}$.
These maximal bond dimension $\chi$ and the time step $\delta t$ were carefully chosen by
\begin{itemize}
\item[(i)] benchmarking the imbalance computed by TEBD against the numerically-exact integration of the dynamics generated by Eq.~(1) in the main manuscript for systems of $L=8$ sites, see Fig.~\ref{fig:MPS_benchmark1},
\item[(ii)] carefully checking the convergence of the TEBD-MPDO dynamics with respect to the bond dimension $\chi$ and $\delta t$ for larger systems of $L=32$ sites,  see Fig.~\ref{fig:MPS_benchmark2}.
\end{itemize}
Figure~(\ref{fig:MPS_benchmark1}) shows that, for strong enough disorder strength $h=20$ and for two representative values of the quantum jump fugacity $\zeta$, the error made with TEBD reduces by increasing the bond dimension $\chi$ and by reducing the Trotter time step $\delta t$.
Errors are kept below $10^{-5}$ with $\chi\geq 2^{6}$ and $\delta t \leq 10^{-2}$.
Figure~(\ref{fig:MPS_benchmark2}) shows that for strong enough disorder $h=20$, the variations of the imbalance computed with TEBD when doubling the bond dimension are smaller than $10^{-5}$  for $\chi \geq 2^6$ and $\delta t\leq10^{-2}$ in the case $\zeta=0.2$. The $\zeta=1.0$ case requires $\chi \geq 2^7$ and $\delta t\leq10^{-2}$. Altogether, this demonstrates that the imbalance dynamics presented in Fig.~6 of the main manuscript are properly converged.

We also numerically observed that, for smaller disorder strengths, the convergence of the imbalance dynamics requires larger bond dimensions, indicating that steady states of the Liouvillian closer to the delocalized regime have a larger operator entanglement entropy~\cite{Wellnitz2022rise}.
This is analogous to the case of closed many-body localized systems, where the entanglement growth is faster closer to the thermal phase and hence requires larger bond dimensions~\cite{Znidaric2018Entanglement,Elmer2021Many}.
\bibliography{references.bib}